\documentclass[IEEEjournal]{IEEEtran}
\usepackage{graphicx}
\usepackage{subfigure} 
\usepackage{algorithm}
\usepackage{algorithmic}
\usepackage{cite}
\usepackage{amsfonts}
\usepackage{amssymb} 
\usepackage{psfrag}
\usepackage{footnote}
\usepackage{amsmath} 
\usepackage[switch,pagewise]{lineno}
\usepackage{stfloats}
\usepackage{rotating}
\usepackage{url}
\usepackage{multirow} 
\usepackage{enumerate}
\usepackage{amsthm} 

\usepackage{accents}

\usepackage{bm}

\makeatletter
\def\ubar{\underaccent{{\underline{\mskip7mu}}}}
\def\lubar[#1]{\underaccent{{\underline{\mskip#1mu}}}}
\makeatother

\begin{document}

\title{AWG-based Nonblocking Shuffle-Exchange Networks}
\author{
        Tong~Ye,~\IEEEmembership{Member,~IEEE,}
        Jingjie~Ding,
        Tony~T.~Lee,~\IEEEmembership{Fellow,~IEEE,}
        and~Guido~Maier,~\IEEEmembership{Senior~Member,~IEEE}

\thanks{This work was supported by the National Science Foundation of China (61571288, 61671286, and 61433009).}
\thanks{Tong Ye, Jingjie Ding are with the State Key Laboratory of Advanced Optical Communication Systems and Networks, Shanghai Jiao Tong University, Shanghai 200240, China. (email: \{yetong,mrdingjay\}@sjtu.edu.cn).}
\thanks{Tony T. Lee is with the Chinese University of Hong Kong (Shenzhen), Shenzhen 518172, China (e-mail: tonylee@cuhk.edu.cn). }
\thanks{Guido Maier is with the Dipartimento di Elettronica, Informazione e Bioingegneria, Politecnico di Milano, Milan 20133, Italy (e-mail: guido.maier@polimi.it).}
}

%
\maketitle


\begin{abstract}
Optical shuffle-exchange networks (SENs) have wide application in different kinds of interconnection networks. This paper proposes an approach to construct modular optical SENs, using a set of arrayed waveguide gratings (AWGs) and tunable wavelength converters (TWCs). According to the wavelength routing property of AWGs, we demonstrate for the first time that an AWG is functionally equivalent to a classical shuffle network by nature. Based on this result, we devise a systematic method to design a large-scale wavelength-division-multiplexing (WDM) shuffle network using a set of small-size AWGs associated with the same wavelength set. Combining the AWG-based WDM shuffle networks and the TWCs with small conversion range, we finally obtain an AWG-based WDM SEN, which not only is scalable in several ways, but also can achieve 100\% utilization when the input wavelength channels are all busy. We also study the routing and wavelength assignment (RWA) problem of the AWG-based WDM SEN, and prove that the self-routing property and the nonblocking routing conditions of classical SENs are preserved in such AWG-based WDM SEN.
\end{abstract}


\IEEEpeerreviewmaketitle
\section{Introduction}\label{section:intro}

\IEEEPARstart{W}{ith} the coming of big data era, high-speed switching networks become more and more important. Currently, massive data exchange is very common within the information systems at different scales. On a multi-processor system, different processors work together through a switching network to provide high-power data processing capability \cite{Mehrnaz:JON2013,Mashhadi:icIVPR2017,Sharifi:PAAP2008}. This is also true for mega data centers, where more than $10^{5}$ servers \cite{Lugones:FGCS2014} are interconnected via a large-scale switching network to deliver high-quality cloud computing service or high-performance computation service. Also, in an even larger area, high-capacity switching network is now indispensable to big data transfers among different data centers \cite{Ueda:OFC2017}. For example, it is reported that the traffic rate per link in a data center has come to tens of Tb/s \cite{Ueda:OFC2017,Park:IMISUC2017}, and will reach Pb/s in the near future \cite{Lugones:FGCS2014,Ueda:OFC2017,Miao:JLT2016}.

In the meanwhile, optical shuffle-exchange networks (SENs) have exhibited several advantages in high-speed interconnection \cite{Veselovsky:DICTAP2012}. First, as a kind of optical switches, an optical SEN can provide large switching capacity \cite{Lugones:FGCS2014,Othman:TTT10,Shahida:ISHCNET2008,Arya:NGCT2015}, especially when it employs wavelength division multiplexing (WDM) technology \cite{Kamchevska:JLT2016}. Second, the structure of SENs is very regular and thus easy to deploy \cite{Lugones:FGCS2014,Wong:AO1995}. Third, the SEN has a less network diameter than other networks, and thus less implementation cost \cite{Sharifi:PAAP2008}. Fourth, the routing algorithm of the SEN is simple since it has a self-routing property \cite{Sunil:AASPAA1995,TTlee:Book2010}: the path between an input-output pair is uniquely determined by the addresses of the input and the output. Hence, optical SEN has been considered as a promising candidate to provide high-speed interconnection for different network application scenarios \cite{Mehrnaz:JON2013,Mashhadi:icIVPR2017,Lugones:FGCS2014,Veselovsky:DICTAP2012,Abdullah:ICCI2006}.

\begin{figure}[t]
\centering
\includegraphics[scale=0.86]{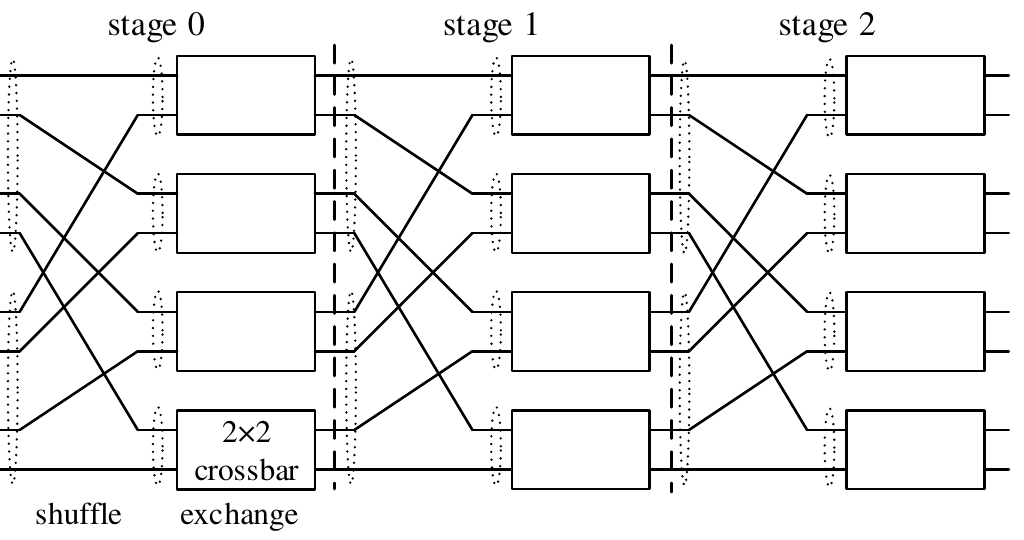}
\caption{An $8\times8$ shuffle exchange network.}
\label{fig1}
\end{figure}

The SEN is a kind of multi-stage switching networks, as Fig. \ref{fig1} illustrates. An ${m^n}\times{m^n}$ SEN consists of $n$ cascaded switching stages, each of which is a shuffle network followed by an exchange network. For example, the SEN in Fig. \ref{fig1} is a SEN with $m=2$ and $n=3$. Each shuffle network is an $m^n\times m^n$ perfect shuffle. It splits its inputs and its outputs into $m$ input groups and $m^{n-1}$ output groups, respectively, and connects an input group and an output group via one and only one link. Each exchange network contains $m^{n-1}$ $m\times m$ crossbars, and each crossbar is attached to an output group of the shuffle network. In other words, each crossbar performs switching function for an output group. For example, each shuffle network in Fig. \ref{fig1} is a $2^3\times 2^3$ perfect shuffle, and it partitions 8 inputs and 8 outputs into $m=2$ input groups and $m^{n-1}=2^{3-1}=4$ output groups, each of which is encircled by a dotted circle. Each exchange network in Fig. \ref{fig1} consists of $m^{n-1}=2^{3-1}=4$ $2\times 2$ crossbars, and the inputs of a crossbar are actually the outputs of an output group of a shuffle network.

Though electrical SEN is already a mature technology, the design of scalable optical SENs remains a big challenge. Up to now, several kinds of optical components, such as micro-electromechanical system (MEMS) \cite{Lugones:FGCS2014}, arrayed waveguide grating (AWG) \cite{Maier:ICCCN2011}, and tunable wavelength converter (TWC) \cite{Liu:JLT2007}, have been available to construct optical switches. The power consumption of optical components typically increases with their size, which, however, cannot be very large, due to cost, synthesis difficulty, or physical-layer performance. For example, an AWG with a large port count induces serious coherent crosstalk when the same wavelength is fed into a number of inputs \cite{Gaudino:ICC2008}. Such coherent crosstalk causes signal degradation and is very hard to get rid of. Another example is TWC with large wavelength conversion range, which is powerful but expensive \cite{Pattavina:INFOCOM2006}. Also, the number of wavelengths required by the network, referred to as wavelength granularity in this paper, should not be very large, since the number of wavelengths available in optical transmission window is limited \cite{Weichenberg:JOCN2009}. Furthermore, the number of links in each shuffle network should be carefully considered. It is well-known that the cabling complexity has become one of the development bottle-neck of high-speed switching nodes \cite{Zong:OFC2016}. At last, the optical SEN should be compact and easy for maintenance, which is also very important for practical applications. Though there exist several optical designs, they cannot meet all these requirements at the same time.

\subsection{Previous work}
In the early phase, the design of optical SENs was mainly based on free-space optics (FSO) technology. Ref. \cite{Brenner:AO1988,Stirk:AO1988,YANG:OSN2014} proposed several designs, employing customized complex lens, polarizing beam splitters and micro-blazed grating arrays. In \cite{Waterson:AO1994,Cao:AO1992,Wong:AO1995}, different kinds of exchange networks were implemented with spatial light modulators, ferroelectric liquid crystals, calcite crystals, and arrayed optical switches, respectively. However, these FSO devices are typically costly, bulky, and difficult to adjust and maintain \cite{Son:DCN2017}. Therefore, the FSO-based schemes are not suitable for most of practical applications.

In \cite{Othman:TTT10} and \cite{Yang:TPDS2001}, two kinds of fiber-based SENs were proposed to avoid the disadvantages of the FSO-based SEN. In such SENs, optical fiber was used to construct the optical shuffle networks, while the MEMSs \cite{Yang:TPDS2001} or the electric-optical switches \cite{Othman:TTT10} were employed to implement the optical exchange networks. One feature of such kind of SENs is that each fiber only carries one optical signal. Thus, there are $N$ fibers in each optical shuffle network if the port count of the SEN is $N$. Clearly, the cabling complexity of optical shuffle networks in the SEN will be high if $N$ is large.

Recently, combination of AWGs and TWCs provides a new way to construct SENs \cite{Maier:ICCCN2011}. An $N\times N$ AWG is an $N\times N$ passive optical component \cite{Maier:ICCCN2011}, each port of which carries the same set of $N$ wavelengths. The function of AWGs is to forward the signal from an input to an output without any contention. On the other hand, the TWC can convert an input wavelength to any of the output wavelengths in the conversion range. With the TWCs at each input, the AWG can perform high-speed switching function \cite{Maier:ICCCN2011,Kachris:MCOM2013,Ye:TON2015,Ge:TON2017,Lucerna:HPSR2011,Testa:book2018,Sato:MCOM2013}. In particular, a TWC can send an optical packet from an input to an output of the AWG, if it converts the incoming wavelength to one of $N$ wavelengths. Based on such feature, Ref. \cite{Maier:ICCCN2011} proposed an $N\times N$ AWG-based SEN with $\log_2N$ cascaded switching stages, each of which was an $N\times N$ AWG with each output attached by a TWC. However, the design in \cite{Maier:ICCCN2011} has several drawbacks as follows. Firstly, in order to eliminate coherent crosstalk, the AWG-based SEN in \cite{Maier:ICCCN2011} does not make full use of the WDM property of AWGs. Only $N$ wavelength channels of each AWG are employed, and thus the AWG utilization is only $1/N$. Secondly, when $N$ is large, the scalability of this design is not good due to the following reasons: 1) it needs large-scale AWGs, which results in large wavelength granularity; 2) the conversion range of the employed TWCs is large; 3) a complex interconnection between the TWCs and the AWG at each stage is required to avoid the coherent crosstalk induced by the large-scale AWG.

\subsection{Our Work}
The focus of this paper is on the construction of AWG-based WDM SENs. Different from that in \cite{Maier:ICCCN2011}, the design of this paper takes full advantage of the WDM property of AWGs, such that the above-mentioned drawbacks can be avoided.

To achieve this goal, the important step is to construct a modular AWG-based WDM shuffle network, in which the wavelength channels of each AWG are fully utilized. We demonstrate that an AWG is functionally equivalent to a shuffle network by nature if the wavelength channels at each port are considered as a channel group. Based on this result, we devise a systematic method to construct a large-scale WDM shuffle network using a collection of small-size AWGs associated with the same wavelength set. We show that the cabling complexity of such modular AWG-based shuffle networks is remarkably cut down and serious coherent crosstalk is suppressed.

We then propose an AWG-based WDM SEN by combining AWG-based WDM shuffle networks and TWC modules, each of which is constructed from a set of TWCs. We show that the conversion range of TWCs in the network is reduced since the size of each employed AWG is small. We also study the routing and wavelength assignment (RWA) problem of the AWG-based WDM SEN. We show that the self-routing property and the nonblocking routing conditions of the classical SEN are also preserved in such AWG-based WDM SEN.

In summary, our main contribution includes:
\begin{enumerate}[{1)}]
\item
We find, for the first time, the functional equivalence between a single AWG and a shuffle network;
\item
We develop a systematic method to construct modular AWG-based shuffle networks, of which the AWG size, the wavelength granularity, and the cabling complexity are scaled down;
\item
We design AWG-based WDM SENs, of which the coherent crosstalk is small and the utilization of the optical components is 100\% if the input channels are all busy.
\end{enumerate}

The rest of this paper is organized as follows. We first demonstrate the functional equivalence between an AWG and a shuffle network in Section \ref{section:shuffle}, and then we develop a systematic approach to construct a modular AWG-based WDM shuffle network in Section \ref{section:modular}. Based on the result in Section \ref{section:modular}, we design  AWG-based WDM SENs in Section \ref{section:sen}. At last, Section \ref{section:rwa} shows that the self-routing property and the nonblocking conditions of classical SENs are also preserved by such WDM SENs. Finally, Section \ref{section:conclu} concludes this paper.

\section{Generalized AWG-Based Shuffle Network}\label{section:shuffle}
AWG is a kind of passive wavelength router, which is able to provide exact one connection between each input and each output. Because of such unique feature, the AWG has been used to provide broadband optical interconnections for different applications, such as shuffle network in \cite{Maier:ICCCN2011}. However, the efficient way for AWGs to fulfill the function of shuffle networks is still unknown.

In this section, we show for the first time that an AWG is functionally equivalent to a shuffle network. In particular, we first recall the definition of shuffle networks in Section \ref{section:shuffle-a}, and then demonstrate why and how the AWG can be equivalent to a shuffle network in Section \ref{section:shuffle-b}.

\subsection{Generalized Shuffle Network} \label{section:shuffle-a}
\begin{figure}[t]
\centering
\includegraphics[scale=1]{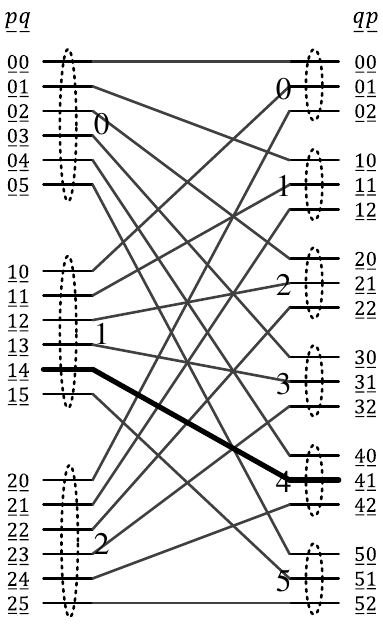}
\caption{An example of $18\times 18$ generalized shuffle $\mathcal{N}(3,6)$.}
\label{fig2}
\end{figure}

Consider an $N\times N$ interconnection network with $N$ inputs and $N$ outputs. Assume that $N$ inputs can be equally divided into $m$ groups, each with $l$ ports, and $N$ outputs can be partitioned into $l$ groups, each with $m$ ports, where $N=ml$. The $N\times N$ generalized shuffle network can be defined as follows.
\newtheorem{definition}{Definition}
\begin{definition}
An $N\times N$ interconnection network is an $N\times N$ generalized shuffle network, denoted by $\mathcal{N}(m,l)$, if the $q$th port of the $p$th input group connects to the $p$th port of the $q$th output group, where $p=0,1,\cdots,m-1$ and $q=0,1,\cdots,l-1$.
\end{definition}

Fig. \ref{fig2} plots an $18\times 18$ shuffle network $\mathcal{N}(3,6)$, where 18 inputs and 18 outputs are evenly divided into 3 groups and 6 groups, respectively. Also, the 5th input of input group 0 connects with the 0th output of output group 5.

We assign each port a two-field address, of which the first field is called group field while the second field is port field. In particular, we assign a two-field address $\ubar{p} \ubar{q}$ to the $q$th input of the $p$th input group, and a two-field address $\ubar{q} \ubar{p}$ to the $p$th output of the $q$th output group, where the underlined numbers in the field address are used to represent the corresponding port and/or group sub-addresses. Under such numbering scheme, input $\ubar{p} \ubar{q}$ connects with output $\ubar{q}\ubar{p}$. We thus denote this connection by $C(\ubar{p}\ubar{q},\ubar{q}\ubar{p})$. For example, input $\ubar{1}\ubar{4}$ connects with output $\ubar{4}\ubar{1}$ via connection $C(\ubar{1}\ubar{4},\ubar{4}\ubar{1})$ in Fig. \ref{fig2}.

From the above definition, it is easy to show that shuffle networks have the following two features:
\begin{enumerate}[{F1.}]
\item
An input group connects with an output group via exact one connection;
\item
An input connects to an output, whose address is formed by exchanging the two sub-addresses of the input address.
\end{enumerate}
An $N\times N$ interconnection network is functionally equivalent to shuffle network $\mathcal{N}(m,l)$ if it can fulfill these two connection features.

\subsection{Single-AWG based Shuffle Network} \label{section:shuffle-b}
\begin{figure*}
\centering
\subfigure[]{\includegraphics[scale=0.89]{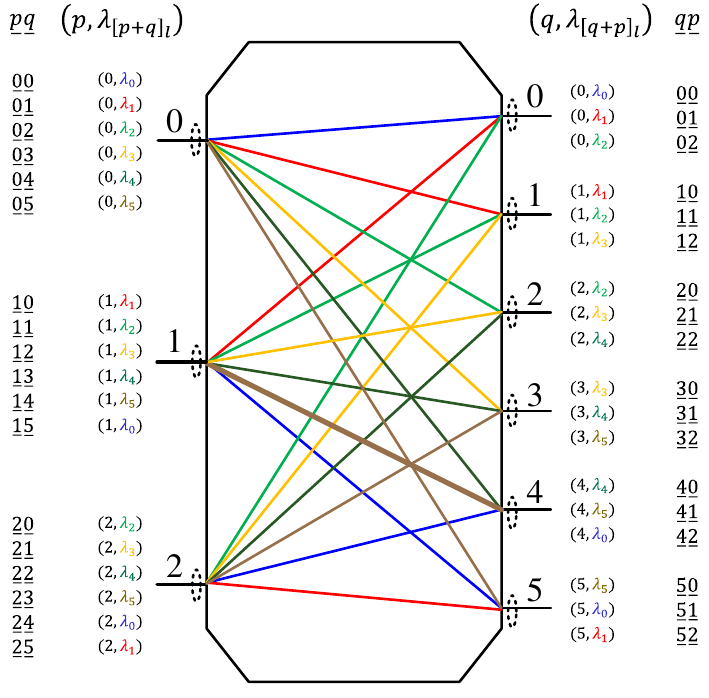}\label{fig3a}}
\ \ \ \ \ \ \ \ \ \ \ \ \ \
\subfigure[]{\includegraphics[scale=0.90]{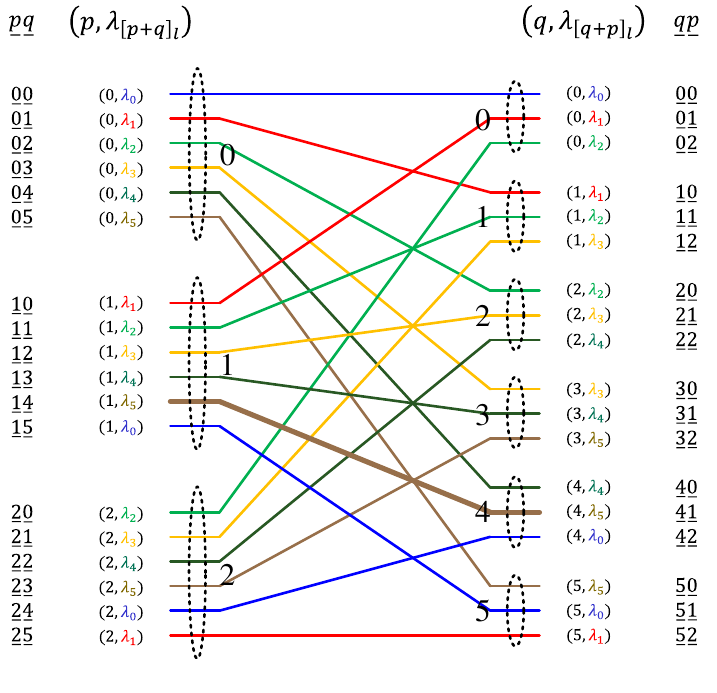}\label{fig3b}}
\caption{Equivalence between a single AWG and a shuffle: (a) AWG $\mathcal{A}(3,6)$ and (b) its space representation.}
\end{figure*}

An $m\times l$ AWG, denoted as $\mathcal{A}(m,l)$, has $m$ inputs and $l$ outputs, which are labelled from top to bottom. AWG $\mathcal{A}(m,l)$ is associated with a wavelength set $\Lambda=\{\lambda_0,\lambda_1,\cdots,\lambda_{|\Lambda |-1}\}$ in a free spectrum range (FSR), where $|\Lambda|=\mbox{max}\{m,l\}$. Without loss of generality, we assume that $l>m$ in this section, and thus $|\Lambda|=l$. This paper only considers the wavelength channels in the main FSR, because AWGs suffer physical performance degradation at the wavelength channels outside the main FSR \cite{Kamei:JLT2009}. In the main FSR, each input of AWG $\mathcal{A}(m,l)$ carries $l$ wavelength channels, and each output has $m$ wavelength channels. Fig. \ref{fig3a} gives an example of a $3\times 6$ AWG $\mathcal{A}(3,6)$.

The AWG is a passive wavelength router. In an $m\times l$ AWG, input $p$ is connected to output $q$ via wavelength $\lambda_i$, where:

\begin{equation}{
i=[p+q]_l
\label{eq:awg1}
}\end{equation}
or
\begin{equation}{
q=[i-p]_l
\label{eq:awg2}
}\end{equation}
or
\begin{equation}{
p=[i-q]_l,
\label{eq:awg3}
}\end{equation}where $[X]_l \triangleq (X \mbox{ mod } l)$. The wavelength routing property (\ref{eq:awg1}) clearly shows that input $p$ and output $q$ are connected by exact one wavelength channel $\lambda_{[p+q]_l}$, of which the connection is denoted by $\tilde{C}_A(p,q,\lambda_{[p+q]_l})$. Such connection feature is consistent with the feature F1 of shuffle network $\mathcal{N}(m,l)$.

\begin{table}
  \centering
  \caption{Wavelength Routing Table $\bm{T}_{A}$ of a $3\times 6$ AWG $\mathcal{A}(3,6)$}\label{table:1awg}
  \begin{tabular}{|c|c|c|c|c|c|c|}
     \hline
     \bm{$p\backslash q$} & \bm{$0$} & \bm{$1$} & \bm{$2$} & \bm{$3$} & \bm{$4$} & \bm{$5$} \\
     \hline
     \bm{$0$} & $\lambda_0$ & $\lambda_1$ & $\lambda_2$ & $\lambda_3$ & $\lambda_4$ & $\lambda_5$ \\
     \hline
     \bm{$1$} & $\lambda_1$ & $\lambda_2$ & $\lambda_3$ & $\lambda_4$ & $\lambda_5$ & $\lambda_0$ \\
     \hline
     \bm{$2$} & $\lambda_2$ & $\lambda_3$ & $\lambda_4$ & $\lambda_5$ & $\lambda_0$ & $\lambda_1$ \\
     \hline
   \end{tabular}
\end{table}

Also, the wavelength routing property (\ref{eq:awg1}) of $\mathcal{A}(m,l)$ can be described by an $m\times l$ routing table, denoted by $\bm{T}_{A}$, where the $p$th row and the $q$th column are corresponding to input $p$ and output $q$, and the entry at the intersection, denoted by $(p,q)$, records the wavelength channel $\lambda_{[p+q]_l}$ of connection $\tilde{C}_A(p,q,\lambda_{[p+q]_l})$. For example, Table \ref{table:1awg} is the routing table of AWG $\mathcal{A}(3,6)$ in Fig. \ref{fig3a}, and entry $(1,4)$ is the wavelength channel $\lambda_{[p+q]_l}=\lambda_{[1+4]_6}=\lambda_5$ of $\tilde{C}(1,4,\lambda_5)$. Therefore, the entries at row $p$ (column $q$) are corresponding to the wavelength channels at input $p$ (output $q$).

According to the structure of $\bm{T}_{A}$, we assign addresses to the input channels and the output channels as follows:
\begin{enumerate}[{1)}]
\item
Input channel: Assign addresses to the wavelength channels at row $p$ (i.e., input group $p$) from left to right. The $q$th entry at row $p$ is assigned with field address $\ubar{p} \ubar{q}$ and two-tuple address $(p,\lambda_{[p+q]_l})$.
\item
Output channel: Assign addresses to the wavelength channels at column $q$ (i.e., output group $q$) from top to bottom. The $p$th entry at column $q$ is assigned with field address $\ubar{q}\ubar{p}$ and two-tuple address $(q,\lambda_{[q+p]_l})$.
\end{enumerate}
In the field address, the first field is called port field, and the second field is referred to as channel field.

With such numbering scheme, entry $(p,q)$ in $\bm{T}_A$ delineates both input channel $\ubar{p}\ubar{q}$ and output channel $\ubar{q}\ubar{p}$, which means they connect to each other through $\lambda_{[p+q]_l}$. To describe such kind of connectivity explicitly, we replace entry $(p,q)$ in $\bm{T}_A$ with input channel $\ubar{p}\ubar{q}$ and output channel $\ubar{q}\ubar{p}$, and obtain a new table, called connectivity table and denoted by $\bm{T}_B$. For example, Table \ref{table:cofawg} gives the connectivity table of $\mathcal{A}(3,6)$, in which entry $(1,4)$ shows that input channel $\ubar{1}\ubar{4}$ connects with output channel $\ubar{4}\ubar{1}$. This property is quite similar to the connection feature F2 of shuffle network $\mathcal{N}(m,l)$.

\begin{table}
  \centering
  \caption{Connectivity Table $\bm{T}_B$ of $\mathcal{A}(3,6)$}\label{table:cofawg}
  \begin{tabular}{|c|c|c|c|c|c|c|}
     \hline
     \bm{$p\backslash q$} & \bm{$0$} & \bm{$1$} & \bm{$2$} & \bm{$3$} & \bm{$4$} & \bm{$5$} \\
     \hline
     \bm{$0$} & $\ubar{0}\ubar{0},\ubar{0}\ubar{0}$ & $\ubar{0}\ubar{1},\ubar{1}\ubar{0}$ & $\ubar{0}\ubar{2},\ubar{2}\ubar{0}$ & $\ubar{0}\ubar{3},\ubar{3}\ubar{0}$ & $\ubar{0}\ubar{4},\ubar{4}\ubar{0}$ & $\ubar{0}\ubar{5},\ubar{5}\ubar{0}$ \\
     \hline
     \bm{$1$} & $\ubar{1}\ubar{0},\ubar{0}\ubar{1}$ & $\ubar{1}\ubar{1},\ubar{1}\ubar{1}$ & $\ubar{1}\ubar{2},\ubar{2}\ubar{1}$ & $\ubar{1}\ubar{3},\ubar{3}\ubar{1}$ & $\ubar{1}\ubar{4},\ubar{4}\ubar{1}$ & $\ubar{1}\ubar{5},\ubar{5}\ubar{1}$ \\
     \hline
     \bm{$2$} & $\ubar{2}\ubar{0},\ubar{0}\ubar{2}$ & $\ubar{2}\ubar{1},\ubar{1}\ubar{2}$ & $\ubar{2}\ubar{2},\ubar{2}\ubar{2}$ & $\ubar{2}\ubar{3},\ubar{3}\ubar{2}$ & $\ubar{2}\ubar{4},\ubar{4}\ubar{2}$ & $\ubar{2}\ubar{5},\ubar{5}\ubar{2}$ \\
     \hline
   \end{tabular}
\end{table}

Therefore, if the channels at one port are regarded as a channel group, $\mathcal{A}(m,l)$ and $\mathcal{N}(m,l)$ have the following equivalence:
\newtheorem{property}{Property}
\begin{property}
A single AWG $\mathcal{A}(m,l)$ is equivalent to a shuffle network $\mathcal{N}(m,l)$ in terms of the connectivity between input channels and output channels.
\label{prop1}
\end{property}

\begin{IEEEproof}
This property holds due to the fact that each input-output pair of $\mathcal{A}(m,l)$ is connected via only one wavelength channel, which is elaborated as follows:
\begin{enumerate}[{1)}]
\item
In $\mathcal{A}(m,l)$, input $p$ connects to output $q$ via only one connection $\tilde{C}_A(p,q,\lambda_{[p+q]_l})$. It follows that there exists an one-one and onto mapping

\begin{equation}{
p,q \leftrightarrow \tilde{C}_A(p,q,\lambda_{[p+q]_l}).
\label{eq:p1proof1}
}\end{equation}
\item
According to the property of shuffle network $\mathcal{N}(m,l)$, input group $p$ is connected to output group $q$ via a unique connection $C(\ubar{p}\ubar{q},\ubar{q}\ubar{p})$. Thus, there is an one-one and onto mapping

\begin{equation}{
p,q \leftrightarrow C(\ubar{p}\ubar{q},\ubar{q}\ubar{p}).
\label{eq:p1proof2}
}\end{equation}
\end{enumerate}
Thus, if each port of $\mathcal{A}(m,l)$ is regarded as a channel group, there is an one-one and onto mapping between connection $\tilde{C}_A(p,q,\lambda_{[p+q]_l})$ in $\mathcal{A}(m,l)$ and connection $C(\ubar{p}\ubar{q},\ubar{q}\ubar{p})$ in $\mathcal{N}(m,l)$, i.e.,

\begin{equation}{
\tilde{C}_A(p,q,\lambda_{[p+q]_{l}}) \leftrightarrow C(\ubar{p}\ubar{q},\ubar{q}\ubar{p}).
\label{eq:p1proof3}
}\end{equation}
This establishes Property \ref{prop1}.
\end{IEEEproof}

The space representation of $\mathcal{A}(m,l)$ is plotted in Fig. \ref{fig3b}, where each line in the left side represents an input channel, each line in the right side stands for an output channel, and that in between is a connection connecting a pair of input channel and output channel. Fig. \ref{fig3b} clearly shows the one-one and onto mapping between the connection in $\mathcal{A}(m,l)$ and that in $\mathcal{N}(m,l)$.

According to Property \ref{prop1}, we can use an $m\times l$ AWG to construct an $N\times N$ single-AWG based shuffle network $\mathcal{A}(m,l)$, where $N=ml$. The cabling complexity is only $O(l)$, since we need $m+l$ fibers at the inputs and outputs when we use it. Also, the $m\times l$ AWG contains $N=ml$ wavelength channels, and thus the utilization of the AWG is 100\% if all the input wavelength channels are busy. However, as either the number of inputs $m$ or that of outputs $l$ becomes large, the single-AWG based shuffle network is not scalable and will suffer from difficult synthesis technique, serious crosstalk, and large wavelength granularity, as we mention in Section \ref{section:intro}.

\section{Modular AWG-based Shuffle Network}\label{section:modular}
In this section, we study the method to construct a modular AWG-based shuffle network, which is the key step for the design of AWG-based SENs. In particular, we consider how to devise $\mathcal{N}(m,l)$ using a set of $r$ $m\times m$ AWGs, where $l=rm$ and $r=1,2,3,\cdots$. From Section \ref{section:shuffle}, we establish the equivalence between a single AWG and a shuffle network through investigating the routing table of AWGs. Thus, Section \ref{modular-a} starts the construction from the design of the routing table of the modular AWG-based shuffle network, based on which we come up with a systematic method to achieve the construction in Section \ref{modular-b}.

\subsection{Routing Table of $\mathcal{W}(m,rm)$}\label{modular-a}
We denote the modular AWG-based shuffle network to be constructed as $\mathcal{W}(m,rm)$. In $\mathcal{W}(m,rm)$, there are $m\cdot rm=rm^2$ input wavelength channels and $rm^2$ output wavelength channels. These $rm^2$ input channels are divided into $m$ input groups, each with $rm$ input channels, and $rm^2$ output channels are divided into $rm$ output groups, each with $m$ output channels.

Since the building blocks of $\mathcal{W}(m,rm)$ are $m\times m$ AWGs, of which each port carries the same set of $m$ wavelengths $\Lambda=\{\lambda_0,\lambda_1,\cdots,\lambda_{m-1}\}$, we need $r$ $m\times m$ AWGs to construct a $\mathcal{W}(m,rm)$ with $rm^2$ input channels and $rm^2$ output channels. Also, according to the definition of $\mathcal{W}(m,rm)$, every $r$ inputs of $m\times m$ AWGs should be grouped together as an input group, and every output as an output group.

The wavelength routing property of $\mathcal{W}(m,rm)$ can also be described by a wavelength routing table, denoted by $\bm{T}_C$, where each row and each column correspond to an input group and an output group respectively. According to the above description, in general, a legitimate routing table $\bm{T}_C$ should satisfy the following conditions:
\begin{enumerate}[{1)}]
\item
As the wavelength granularity of $\mathcal{A}(m,m)$ is $m$, the routing table can only contain $m$ different wavelengths;
\item
As a kind of shuffle networks, each entry of this table can contain one and only one wavelength;
\item
Since every $r$ inputs of $\mathcal{A}(m,m)$ make up an input group and every output is an output group, the wavelengths in $\Lambda$ must appear $r$ times in each row, and once in each column.
\end{enumerate}

A routing table $\bm{T}_C$ that meets the above conditions can be constructed from the routing table $\bm{T}_A$ of $\mathcal{A}(m,rm)$. There are $m$ rows and $rm$ columns in $\bm{T}_A$, in which each entry has just one wavelength. In particular, entry $(p,q)$ at the intersection of row $p$ and column $q$ contains wavelength channel $\lambda_{[p+q]_{rm}}$, where $p=0,1,\cdots,m-1$ and $q=0,1,\cdots,rm-1$. If we apply modulo operation to all the numbers in $\bm{T}_A$, the row index $p$ remains the same as $[p]_m=p$, the column-index $q$ changes to $q'=[q]_m$, and $\lambda_{[p+q]_{rm}}$ becomes $\lambda_{[p+q']_m}$ since

\begin{equation}{
\big{[}[p+q]_{rm}\big{]}_{m}=[p+q]_{m}=\big{[}p+[q]_{m}\big{]}_{m}=[p+q']_{m}}.
\label{eq:mode}
\end{equation}

It is easy to see that the column index $q'$ periodically increases from 0 to $m-1$, and there are totally $r$ periods in the table after modulo operations. According to (\ref{eq:mode}), every $m$ columns within a period form a subtable, each of which is a routing table of $\mathcal{A}(m,m)$. Thus, there are $r$ such identical subtables in the table. To distinguish these $r$ subtables, one more field is added to the left side of the column index, say $a$. In particular, if a column is the $q$th column of the table, it will be labelled by $\ubar{a}\ubar{{q'}}$ since it is the $q'$th column within the $a$th period, where $q=am+[q]_m=am+q'$. For example, after performing modulo operations on Table \ref{table:1awg} and relabeling the columns, we obtain Table \ref{table:w36}, which contains 2 routing tables of $\mathcal{A}(3,3)$. It is clear that this new table satisfies three conditions mentioned above, and thus is the desired routing table $\bm{T}_C$ of $\mathcal{W}(m,rm)$.

\begin{table}
  \centering
  \caption{Wavelength Routing Table $\bm{T}_C$ of $\mathcal{W}(3,6)$}\label{table:w36}
  \begin{tabular}{|c|c|c|c||c|c|c|}
     \hline
     $\bm{p}\backslash \ubar{\bm{a}} \ubar{\bm{q}}^{'}$ & $\ubar{\bm{0}} \ubar{\bm{0}}$ & $\ubar{\bm{0}} \ubar{\bm{1}}$ & $\ubar{\bm{0}} \ubar{\bm{2}}$ & $\ubar{\bm{1}} \ubar{\bm{0}}$ & $\ubar{\bm{1}} \ubar{\bm{1}}$ & $\ubar{\bm{1}} \ubar{\bm{2}}$ \\
     \hline
     \bm{$0$} & $\lambda_0$ & $\lambda_1$ & $\lambda_2$ & $\lambda_0$ & $\lambda_1$ & $\lambda_2$ \\
     \hline
     \bm{$1$} & $\lambda_1$ & $\lambda_2$ & $\lambda_0$ & $\lambda_1$ & $\lambda_2$ & $\lambda_0$ \\
     \hline
     \bm{$2$} & $\lambda_2$ & $\lambda_0$ & $\lambda_1$ & $\lambda_2$ & $\lambda_0$ & $\lambda_1$ \\
     \hline
   \end{tabular}
\end{table}

\subsection{Construction of $\mathcal{W}(m,rm)$}\label{modular-b}

\begin{figure}
\centering
\subfigure[]{\includegraphics[scale=0.87]{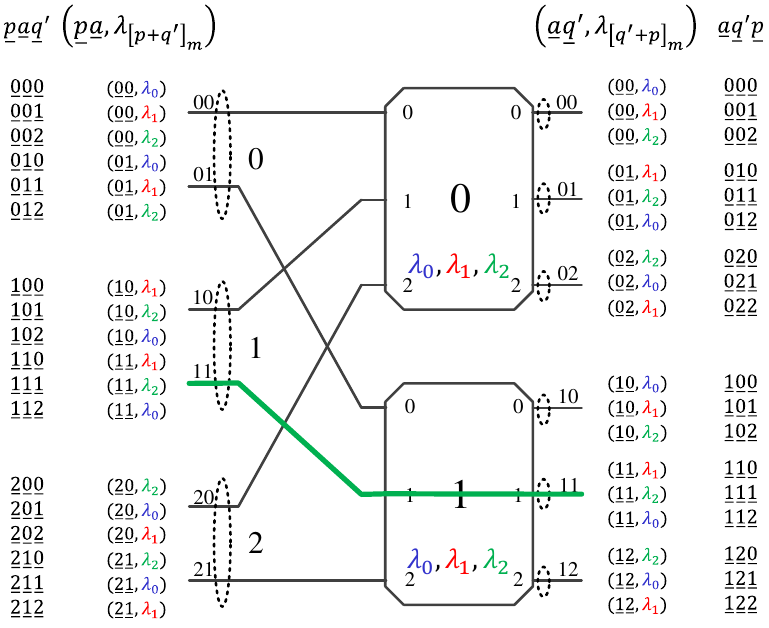}\label{fig4a}}
\ \ \ \ \ \ \ \ \ \ \ \ \
\subfigure[]{\includegraphics[scale=0.86]{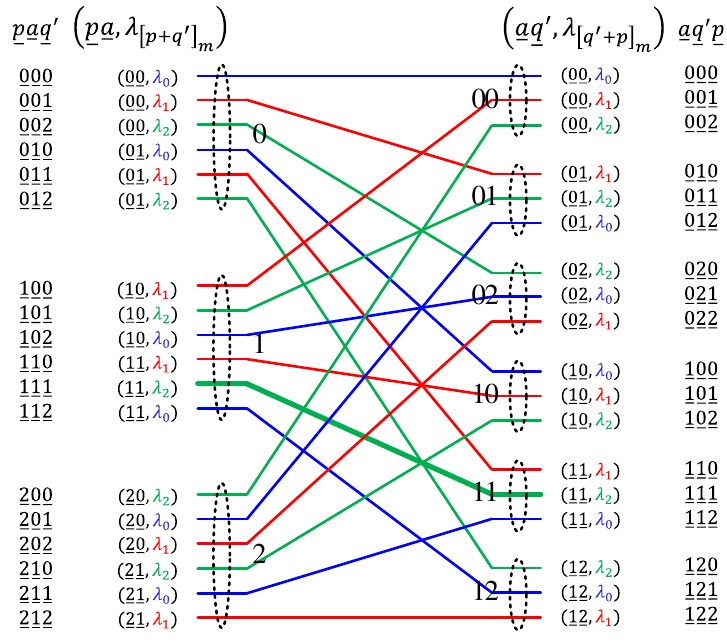}\label{fig4b}}
\caption{Modular AWG-based shuffle network (a) $\mathcal{W}(3,6)$ and (b) its equivalent shuffle network $\mathcal{N}(3,6)$.}
\end{figure}

\begin{table*}
  \centering
  \caption{Connectivity Table $\bm{T}_D$ of $\mathcal{W}(3,6)$}\label{table:cofw36}
  \begin{tabular}{|c|c|c|c||c|c|c|}
     \hline
     $\bm{p}\backslash \ubar{\bm{a}} \ubar{\bm{q}}^{'}$ & $\ubar{\bm{0}} \ubar{\bm{0}}$ & $\ubar{\bm{0}} \ubar{\bm{1}}$ & $\ubar{\bm{0}} \ubar{\bm{2}}$ & $\ubar{\bm{1}} \ubar{\bm{0}}$ & $\ubar{\bm{1}} \ubar{\bm{1}}$ & $\ubar{\bm{1}} \ubar{\bm{2}}$ \\
     \hline
     \bm{$0$} & $\ubar{0}\ubar{0}\ubar{0},\ubar{0}\ubar{0}\ubar{0}$ & $\ubar{0}\ubar{0}\ubar{1},\ubar{0}\ubar{1}\ubar{0}$ & $\ubar{0}\ubar{0}\ubar{2},\ubar{0}\ubar{2}\ubar{0}$ & $\ubar{0}\ubar{1}\ubar{0},\ubar{1}\ubar{0}\ubar{0}$ & $\ubar{0}\ubar{1}\ubar{1},\ubar{1}\ubar{1}\ubar{0}$ & $\ubar{0}\ubar{1}\ubar{2},\ubar{1}\ubar{2}\ubar{0}$ \\
     \hline
     \bm{$1$} & $\ubar{1}\ubar{0}\ubar{0},\ubar{0}\ubar{0}\ubar{1}$ & $\ubar{1}\ubar{0}\ubar{1},\ubar{0}\ubar{1}\ubar{1}$ & $\ubar{1}\ubar{0}\ubar{2},\ubar{0}\ubar{2}\ubar{1}$ & $\ubar{1}\ubar{1}\ubar{0},\ubar{1}\ubar{0}\ubar{1}$ & $\ubar{1}\ubar{1}\ubar{1},\ubar{1}\ubar{1}\ubar{1}$ & $\ubar{1}\ubar{1}\ubar{2},\ubar{1}\ubar{2}\ubar{1}$ \\
     \hline
     \bm{$2$} & $\ubar{2}\ubar{0}\ubar{0},\ubar{0}\ubar{0}\ubar{2}$ & $\ubar{2}\ubar{0}\ubar{1},\ubar{0}\ubar{1}\ubar{2}$ & $\ubar{2}\ubar{0}\ubar{2},\ubar{0}\ubar{2}\ubar{2}$ & $\ubar{2}\ubar{1}\ubar{0},\ubar{1}\ubar{0}\ubar{2}$ & $\ubar{2}\ubar{1}\ubar{1},\ubar{1}\ubar{1}\ubar{2}$ & $\ubar{2}\ubar{1}\ubar{2},\ubar{1}\ubar{2}\ubar{2}$ \\
     \hline
   \end{tabular}
\end{table*}

According to $\bm{T}_C$, we design the $m\times rm$ AWG-based shuffle network $\mathcal{W}(m,rm)$ as follows:
\begin{enumerate}[{S1.}]
\item
For each row of $\bm{T}_C$, create an input group, each of which contains $r$ inputs, and for each column of $\bm{T}_C$, create an output group, each of which contains one output;
\item
Vertically layout $r$ $m\times m$ AWGs, of which the $a$th $m\times m$ AWG is corresponding to the $a$th $m\times m$ subtable of $\bm{T}_C$, where $a=0,1,\cdots,r-1$;
\item
Link the $a$th port of input group $p$, labelled as $\ubar{p}\ubar{a}$, to the $p$th input of the $a$th $m\times m$ AWG, for row $p$ of $\bm{T}_C$ is the $p$th row of the $a$th $m\times m$ subtable of $\bm{T}_C$, where $p=0,1,\cdots,m-1$;
\item
Label output group $q$ by $\ubar{a}\ubar{q}'$ and connect it to the $q'$th output of the $a$th $m\times m$ AWG, if $a=\lfloor q/m \rfloor$ and $q'=[q]_{m}$, where $q=0,1,\cdots,rm-1$.
\end{enumerate}
As Fig. \ref{fig4a} shows, a modular AWG-based shuffle network $\mathcal{W}(3,6)$ is constructed based on $2$ $3\times 3$ AWGs according to Table \ref{table:w36}.

It's easy to check that $\bm{T}_C$ is the routing table of $\mathcal{W}(m,rm)$. We take the connection between input group $p=1$ and output group $\ubar{a}\ubar{q}'= \ubar{1}\ubar{1}$ in Fig. \ref{fig4a} as an example. Output group $\ubar{1}\ubar{1}$ is output $q'=1$ of AWG $a=1$. In order to visit this output group, the connection from input group $p=1$ must pass through input $p=1$ which is linked to input $p=1$ of AWG $a=1$. In other words, this connection goes through the following path:

\begin{small}\begin{eqnarray*}
&& \mbox{input } a=1 \mbox{ of input group } p = 1 \\
&\rightarrow & \text{input } p=1 \mbox{ of AWG } a = 1 \\
&\rightarrow & \mbox{input } q' = 1 \mbox{ of AWG } a = 1.
\label{eq:pathofw}
\end{eqnarray*}\end{small}

According to wavelength routing property (\ref{eq:awg1}), input $p=1$ of AWG $1$ links to output $q'=1$ of AWG $1$ via wavelength $\lambda_i$, where

\begin{equation}{
i=[p+q']_{m}=[1+1]_{3}=2.
\label{eq:modeofw}
}\end{equation}Accordingly, the wavelength in entry $(p,\ubar{a}\ubar{q}')=(1,\ubar{1}\ubar{1})$ is $\lambda_2$ in Table \ref{table:w36}, which verifies that $\bm{T}_C$ is the routing table of $\mathcal{W}(m,rm)$.

As this example demonstrates, there is one and only one connection, denoted as $\tilde{C}(p,\ubar{a}\ubar{q}',\lambda_{[p+q']_m})$, connecting input group $p$ and output group $\ubar{a}\ubar{q}'$. Once input group $p$ and output group $\ubar{a}\ubar{q}'$ are given, the path and the wavelength $\lambda_{[p+q']_m}$ of the connection are uniquely determined: $p$ and $\ubar{a}\ubar{q}'$ decide the path while $p$ and $q'$ give the wavelength. This indicates that no two connections share the same wavelength at the same port, which means $\mathcal{W}(m,rm)$ has the following property:
\begin{property}
$\mathcal{W}(m,rm)$ is wavelength contention-free.
\label{prop2}
\end{property}

\begin{IEEEproof}
Consider two connections $\tilde{C}_{1}(p_{1},\lubar[14]{{a_1}}\lubar[14]{{q'_{1}}},\lambda_{[p_{1}+q'_{1}]_m})$ and $\tilde{C}_{2}(p_{2},\lubar[14]{{a_2}}\lubar[14]{{q'_2}},\lambda_{[p_{2}+q'_{2}]_m})$. Suppose $\tilde{C}_{1}$ and $\tilde{C}_{2}$ originate from the same input and use the same wavelength. Thus, we have

\begin{eqnarray}
p_{1}=p_{2}=p, \nonumber\\
a_{1}=a_{2}=a,
\label{eq:p2proof1}
\end{eqnarray}
and
\begin{equation}{
[p+q'_{1}]_{m} = [p+q'_{2}]_{m}.
\label{eq:p2proof2}
}\end{equation}This indicates that ${q'_{1}=q'_{2}}$ since $q'_{1},q'_{2}=0,1,\cdots,m-1$. Hence, $\tilde{C}_{1}$ and $\tilde{C}_{2}$ are the same connection, which means there is no contention at the input side. Following the same argument, we can show there is no connection at the output side.
\end{IEEEproof}

Property \ref{prop2} shows that all the connections can be set up without any wavelength contention. It follows that, as long as the input channels and output channels are labelled properly, $\mathcal{W}(m,rm)$ can exhibit the same connection features as $\mathcal{N}(m,rm)$.

According to the structure of $\bm{T}_C$, we assign the address to each wavelength channel of $\mathcal{W}(m,rm)$ as follows:
\begin{enumerate}[{1)}]
\item
Input channel: Assign addresses to the channels at row $p$ (i.e., input group $p$) from left to right. The $\ubar{a}\ubar{q}'$th entry at $p$th row is assigned with field address $\ubar{p}\ubar{a}\ubar{q}'$ and two-tuple address $(\ubar{p}\ubar{a},\lambda_{[p+q']_{m}})$.
\item
Output channel: Assign addresses to the channels at column $\ubar{a}\ubar{q}'$ (i.e., output group $\ubar{a}\ubar{q}'$) from top to bottom. The $p$th entry at $\ubar{a}\ubar{q}'$th column is assigned with field address $\ubar{a}\ubar{q}'\ubar{p}$ and two-tuple address $(\ubar{a}\ubar{q}',\lambda_{[q'+p]_{m}})$.
\end{enumerate}

With this numbering scheme, entry $(p,\ubar{a}\ubar{q}')$ in $\bm{T}_C$ actually indicates that input channel $\ubar{p}\ubar{a}\ubar{q}'$ and output channel $\ubar{a}\ubar{q}'\ubar{p}$ are connected together via $\lambda_{[p+q']_{m}}$. Similarly, to express such connectivity more explicitly, we replace $(p,\ubar{a}\ubar{q}')$ with input channel address $\ubar{p}\ubar{a}\ubar{q}'$ and output channel address $\ubar{a}\ubar{q}'\ubar{p}$, and obtain a connectivity table of $\mathcal{W}(m,rm)$, called $\bm{T}_D$. For example, Table \ref{table:cofw36} is the connectivity table of $\mathcal{W}(3,6)$ in Fig. \ref{fig4b}.

Table IV clearly shows that modular AWG-based shuffle network $\mathcal{W}(m,rm)$ proposed in this section and network $\mathcal{N}(m,rm)$ have the following equivalence:
\begin{property}
AWG-based shuffle network $\mathcal{W}(m,rm)$ is equivalent to a shuffle network $\mathcal{N}(m,rm)$ in terms of the connectivity between input channels and output channels.
\label{prop3}
\end{property}

\begin{IEEEproof}
In $\mathcal{W}(m,rm)$, let's consider a connection $\tilde{C}(p,\ubar{a}\ubar{q}',\lambda_{[p+q']_m})$, and assume that $q=am+q'$. As we have shown, there exists a one-one and onto mapping

\begin{equation}{
p,q\leftrightarrow p,\ubar{a}\ubar{q}'\leftrightarrow \tilde{C}(p,\ubar{a}\ubar{q}',\lambda_{[p+q]_m}).
\label{eq:p3proof1}
}\end{equation}
According to the connection feature of shuffle network $\mathcal{N}(m,rm)$, the following one-one and onto mapping is set up
\begin{equation}{
p,\ubar{a}\ubar{q}'\leftrightarrow p,q\leftrightarrow C(\ubar{p}\ubar{q},\ubar{q}\ubar{p}).
\label{eq:p3proof2}
}\end{equation}
It follows that there is a one-one and onto mapping between connection $\tilde{C}(p,\ubar{a}\ubar{q}',\lambda_{[p+q']_m})$ in $\mathcal{W}(m,rm)$ and connection $C(\ubar{p}\ubar{q},\ubar{q}\ubar{p})$ in $\mathcal{N}(m,rm)$, say
\begin{equation}{
\tilde{C}(p,\ubar{a}\ubar{q}',\lambda_{[p+q']_m})\leftrightarrow C(\ubar{p}\ubar{q},\ubar{q}\ubar{p}).
}\end{equation}

Also, according to Property 2, all the connections of $\mathcal{W}(m,rm)$ are achievable without any wavelength contention, which completes the proof.
\end{IEEEproof}

$\mathcal{W}(m,rm)$ exhibits several advantages if it is used to construct an $N\times N$ shuffle network, where $N=rm^{2}$. Firstly, compared to the classical $N\times N$ shuffle network, the cabling complexity is small. In particular, the number of fiber links within $\mathcal{W}(m,rm)$ is $rm=N/m$, while that of the classical shuffle network is $rm^2=N$. Secondly, the utilization of $\mathcal{W}(m,rm)$ is 100\% if all the $rm^2$ input wavelength channels are busy, which is similar to the single-AWG based shuffle network. Thirdly, since $\mathcal{W}(m,rm)$ is constructed from $r$ $m\times m$ AWGs, it possesses good scalability in terms of the following aspects:
\begin{enumerate}[{1)}]
\item
the wavelength granularity $m$ is small;
\item
the coherent crosstalk is small,
\end{enumerate}
if the port count of AWGs $m$ is small. According to the recent experiment results reported in \cite{Xiao:ECOC2018}, the power penalty incurred by the coherent crosstalk of $\mathcal{W}(m,rm)$ can be less than 2 dB if we keep $m\leq32$.

\section{Awg-based WDM Shuffle-Exchange Network}\label{section:sen}
In this section, our goal is to construct an $m^{n}\times m^{n}$ AWG-based WDM SEN. An $m^{n}\times m^{n}$ classical SEN is the combination of $m^{n}\times m^{n}$ shuffle networks and $m\times m$ crossbars, each of which performs switching for an output group of a shuffle network. As Section \ref{section:modular} shows, the AWG-based shuffle network is a WDM shuffle network, in which each port carries multiple wavelength channels. Also, it is known that the TWC can perform wavelength conversion, and a module consisting of a set of TWCs can perform wavelength switching for the optical signals carried by a fiber link \cite{Ye:TON2015}. Therefore, we construct the $m^{n}\times m^{n}$ WDM SEN by the combination of the AWG-based shuffle networks and the TWCs in this section.

\subsection{$m^{n}\times m^{n}$ AWG-based Shuffle Network}
The $m^{n}\times m^{n}$ shuffle network $\mathcal{W}(m,m^{n-1})$ is a shuffle network $\mathcal{W}(m,rm)$ with $r=m^{n-2}$. Since there are $m$ input groups, $m^{n-2}$ inputs in each group, and $m$ wavelength channels at each input in $\mathcal{W}(m,m^{n-1})$, for an input wavelength channel $X=\ubar{p}\ubar{a}\ubar{q}'$, the address $X$ can be relabeled by an $m$-ary $n$-field address $\lubar[19]{{x_n}} \lubar[45]{{x_{n-1}}} \cdots \lubar[19]{{x_2}} \lubar[19]{{x_1}}$ where $x_n$ is corresponding to group address $p$, $\lubar[45]{{x_{n-1}}} \cdots \lubar[19]{{x_2}}$ corresponding to port address $a$, and $x_1$ corresponding to wavelength address $q'$. Accordingly, two-tuple address $\tilde{X}=(\ubar{p}\ubar{a},\lambda_{[p+q']_m})$ can be represented as $(\lubar[19]{{x_n}} \lubar[45]{{x_{n-1}}} \cdots \lubar[19]{{x_2}},\lambda_{[x_{n}+x_{1}]_{m}})$. At the output side, there are $m^{n-1}$ output groups (or outputs), and $m$ wavelength channels at each input. Thus, an output wavelength channel $Y=\ubar{a}\ubar{q}' \ubar{p}$ can also be represented as $\lubar[19]{{y_n}} \lubar[45]{{y_{n-1}}} \cdots \lubar[19]{{y_2}} \lubar[19]{{y_1}}$, where $\lubar[19]{{y_n}} \lubar[45]{{y_{n-1}}} \cdots \lubar[19]{{y_2}}$ is corresponding to group address $\ubar{a}\ubar{q}'$, and $y_1$ corresponding to $p$. Correspondingly, two-tuple address $\tilde{Y}=(\ubar{a}\ubar{q}',\lambda_{[q'+p]_m})$ can be rewritten as $(\lubar[19]{{y_n}} \lubar[45]{{y_{n-1}}} \cdots \lubar[19]{{y_2}},\lambda_{[y_{2}+y_{1}]_{m}})$. According to the discussions in Section \ref{section:modular}, $\mathcal{W}(m,m^{n-1})$ has the following connection property:

\begin{figure}
\centering
\subfigure[]{\includegraphics[scale=0.89]{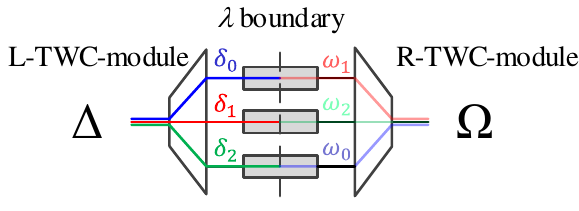}\label{fig5a}}
\subfigure[]{\includegraphics[scale=0.89]{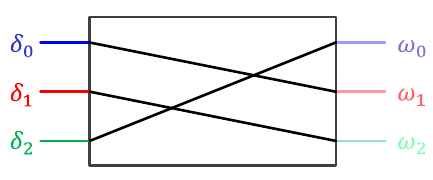}\label{fig5b}}
\caption{Illustration of TWC-modules: (a) a $3\times 3$ TWC-module and (b) its space representation.}
\label{fig5}
\end{figure}

\begin{figure*}[ht]
\centering
\subfigure[]{\includegraphics[scale=0.76]{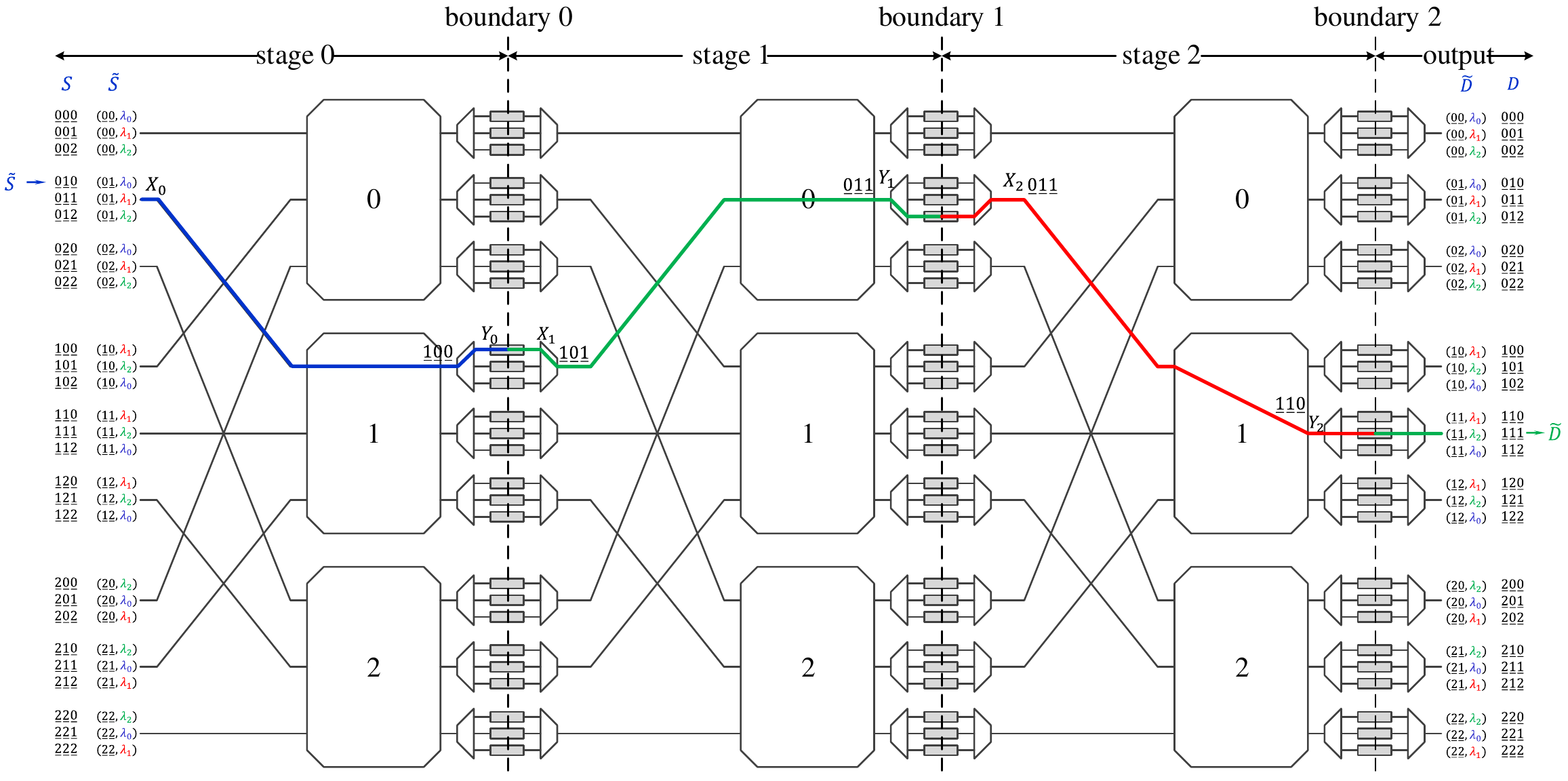}\label{fig6a}}
\subfigure[]{\includegraphics[scale=0.83]{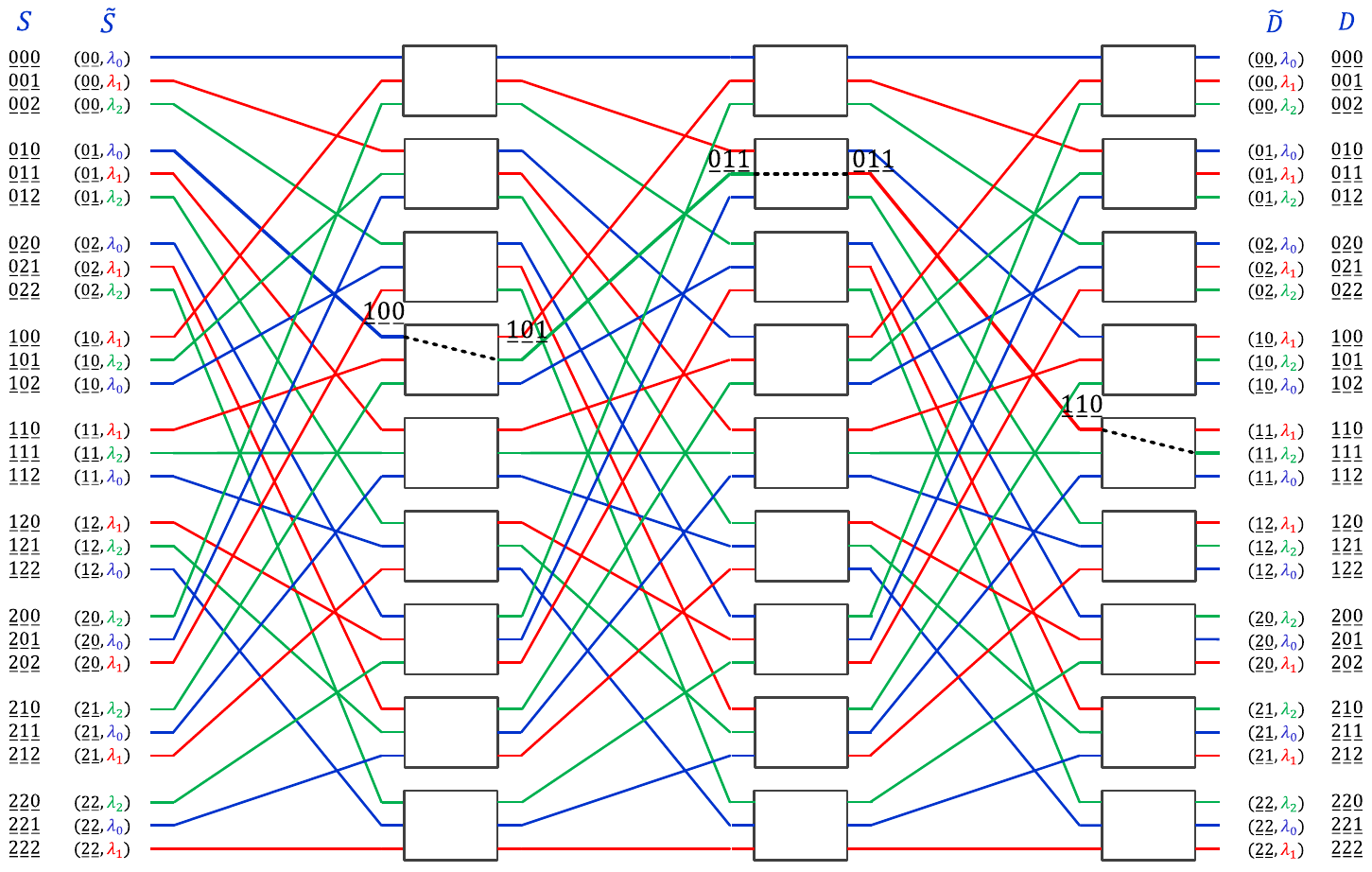}\label{fig6b}}
\caption{Illustration of AWG-based WDM SENs: (a) An AWG-based WDM SEN $\mathcal{S}(m,n)$ and (b) its space representation.}
\label{fig6}
\end{figure*}

\begin{align}
 X=&\ubar{p}\ubar{a}\ubar{q}'=\lubar[19]{{x_n}} \lubar[45]{{x_{n-1}}} \cdots \lubar[19]{{x_2}} \lubar[19]{{x_1}} \nonumber\\
\rightarrow Y=&\ubar{a}\ubar{q}'\ubar{p}=\lubar[45]{{x_{n-1}}} \cdots \lubar[19]{{x_2}} \lubar[19]{{x_1}}\lubar[19]{{x_n}}
\label{eq:conprop1}
\end{align}
or
\begin{align}
\tilde{X}=&(\ubar{p}\ubar{a},\lambda_{[p+q']_m})=(\lubar[19]{{x_n}} \lubar[45]{{x_{n-1}}} \cdots \lubar[19]{{x_2}},\lambda_{[x_{n}+x_{1}]_{m}}) \nonumber\\
 \rightarrow \tilde{Y}=&(\ubar{a}\ubar{q}',\lambda_{[p+q']_m})=(\lubar[45]{{x_{n-1}}} \cdots \lubar[19]{{x_2}}\lubar[19]{{x_1}},\lambda_{[x_{n}+x_{1}]_{m}}).
\label{eq:conprop2}
\end{align}

\subsection{TWC-module}
An $m\times m$ TWC-module consists of $m$ TWCs sandwiched by a $1\times m$ DeMux, and an $m\times 1$ Mux, as Fig. 5(a) illustrates. The function of the TWC-module is to convert a wavelength set $\Delta =\{\delta_0,\delta_1,\cdots,\delta_{m-1}\}$ at the input fiber to a wavelength set $\Omega =\{\omega_0,\omega_1,\cdots,\omega_{m-1}\}$ at the output fiber. In other words, the TWC-module performs a wavelength conversion mapping $\phi :\Delta \rightarrow \Omega$. Thus, if we consider each input wavelength as an input and each output wavelength as an output, an $m\times m$ TWC-module is logically equivalent to an $m\times m$ crossbar. As an example, Fig. \ref{fig5} plots a $3\times 3$ TWC-module and its space representation, which demonstrates the equivalence between the $3\times 3$ TWC-module and the $3\times 3$ crossbar.

The size $m$ of the set $\Omega$ is referred to as conversion range, which actually characterizes the capability of the TWC-module. The conversion range is larger, the synthesis difficulty and thus the cost of the TWC-module are larger. Therefore, the conversion range of the TWC-modules employed in optical switching systems should not be very large.

To facilitate the discussion, we assume that there is a fictitious wavelength boundary in the middle of the TWC-module, and the wavelength conversions happen at this boundary \cite{Ye:TON2015}, as the dashed line in Fig. \ref{fig5a} illustrates. Such wavelength boundary logically divides the TWC-module into left part and right part, called L-TWC-module and R-TWC-module respectively, and the wavelength sets used in these two parts are mutually independent.

\subsection{AWG-based WDM SEN}
\begin{figure*}[!htbp]
\centering
\includegraphics[scale=0.82]{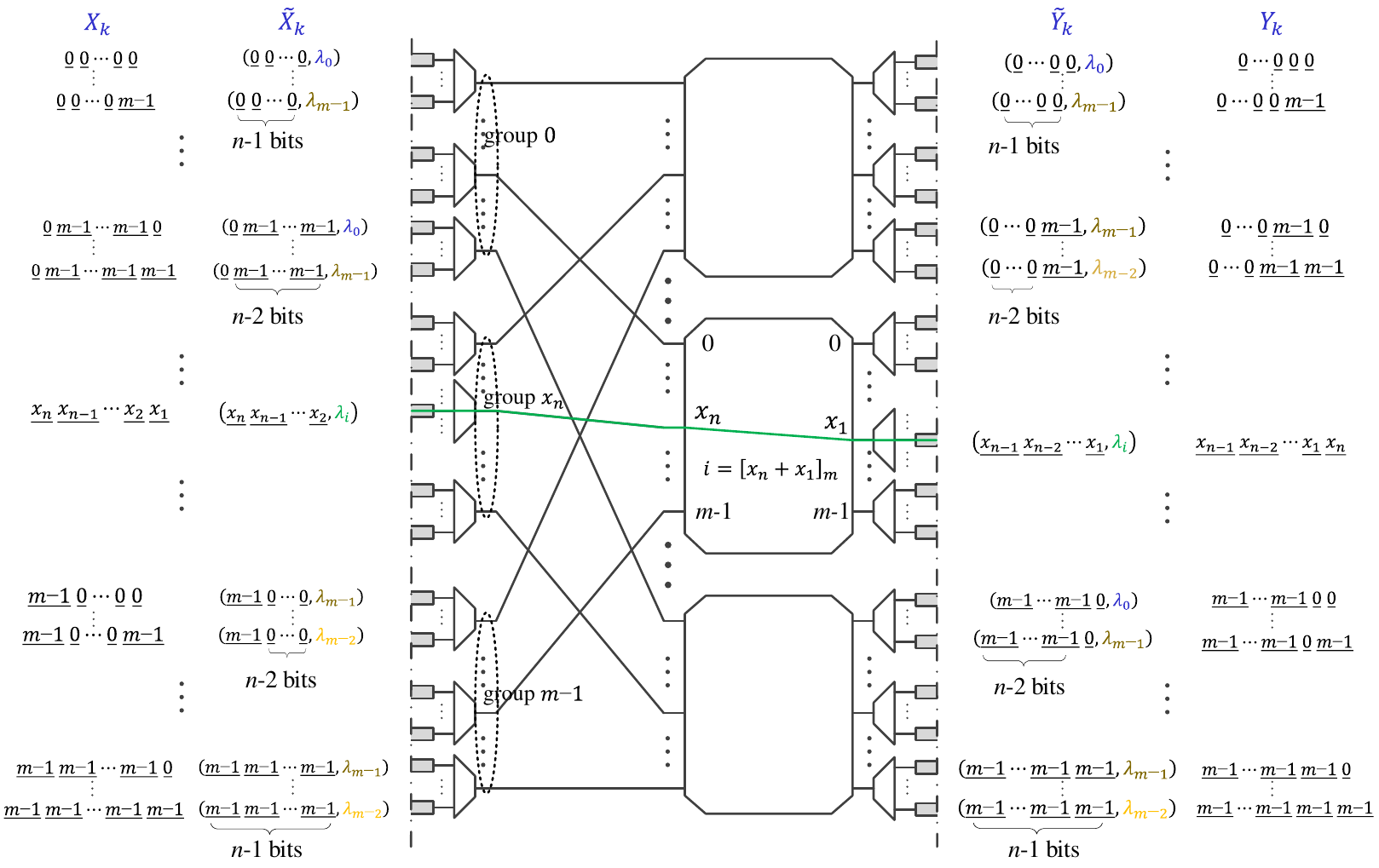}
\caption{The $k$th shuffle stage $\mathcal{W}_k$.}
\label{fig7}
\end{figure*}
An $m^{n}\times m^{n}$ AWG-based WDM SEN proposed in this paper is constructed from $n$ cascaded $m^{n}\times m^{n}$ AWG-based WDM shuffle networks, in each of which every output is attached by an $m\times m$ TWC-module. We denote such AWG-based WDM SEN by $\mathcal{S}(m,n)$. It is clear that there are $n$ columns of $m\times m$ TWC-modules in $\mathcal{S}(m,n)$, and each column contains $m^{n-1}$ TWC-modules. An example is the $3^{3}\times 3^{3}$ AWG-based WDM SEN $\mathcal{S}(3,3)$ in Fig. \ref{fig6a}. This network is functionally equivalent with the $m^{n}\times m^{n}$ classical SEN, which can be demonstrated by the following correspondences:
\begin{enumerate}[{1)}]
\item
Each $m^{n}\times m^{n}$ AWG-based shuffle network $\mathcal{W}(m,m^{n-1})$ is corresponding to a shuffle network $\mathcal{N}(m,m^{n-1})$;
\item
Each $m\times m$ TWC-module is corresponding to an $m\times m$ crossbar.
\end{enumerate}
For example, after the correspondences, the $3^{3}\times 3^{3}$ AWG-based WDM SEN in Fig. \ref{fig6a} changes to the $3^{3}\times 3^{3}$ switching network in Fig. \ref{fig6b}, which is clearly a SEN.

\begin{figure*}[!hb]
\begin{footnotesize}
\hrulefill
\begin{equation}
{\begin{split}
\tilde{S}=(\ubar{0}\ubar{1}&,\lambda_0) \\
S=\ubar{0}&\ubar{1}\ubar{0}
\end{split}}
\xrightarrow[]{\mathcal{W}_0}
{\begin{split}
(\ubar{1}\ubar{0}&,\lambda_0) \\
\ubar{1}&\ubar{0}\ubar{0}
\end{split}}
\xrightarrow[]{\mbox{boundary\ 0}}
{\begin{split}
(\ubar{1}\ubar{0}&,\lambda_2) \\
\ubar{1}&\ubar{0}\ubar{1}
\end{split}}
\xrightarrow[]{\mathcal{W}_1}
{\begin{split}
(\ubar{0}\ubar{1}&,\lambda_2) \\
\ubar{0}&\ubar{1}\ubar{1}
\end{split}}
\xrightarrow[]{\mbox{boundary\ 1}}
{\begin{split}
(\ubar{0}\ubar{1}&,\lambda_1) \\
\ubar{0}&\ubar{1}\ubar{1}
\end{split}}
\xrightarrow[]{\mathcal{W}_2}
{\begin{split}
(\ubar{1}\ubar{1}&,\lambda_1) \\
\ubar{1}&\ubar{1}\ubar{0}
\end{split}}
\xrightarrow[]{\mbox{boundary\ 2}}
{\begin{split}
(\ubar{1}\ubar{1},&\lambda_2)=\tilde{D} \\
\ubar{1}\ubar{1}&\ubar{1}=D
\end{split}}
\label{eq:self-route1}
\end{equation}
\end{footnotesize}
\end{figure*}

As the dotted lines in Fig. \ref{fig6a} illustrates, the middle of each column of TWC-modules in $\mathcal{S}(m,n)$ forms a virtual wavelength boundary, at which the wavelength exchanges are performed. Thus, these $n$ boundaries divide network $\mathcal{S}(m,n)$ into $n+1$ regions, and the wavelength sets in different regions are mutually independent. We refer to the first $n$ regions as shuffle stages, and the last region output stage. Each shuffle stage is a $\mathcal{W}(m,m^{n-1})$ sandwiched by a column of R-TWC modules and a column of L-TWC modules, as Fig. \ref{fig7} demonstrates. Note that the L-TWC module (R-TWC module) actually only performs wavelength multiplexing (demulitplexing) instead of wavelength conversion. Thus, each shuffle stage is actually a shuffle network $\mathcal{W}(m,m^{n-1})$. We thus denote the $k$th shuffle stage as $\mathcal{W}_k$. Let $X_k$ and $Y_k$ be the field addresses of the input wavelength channel and the output wavelength channel, and $\tilde{X}_k$ and $\tilde{Y}_k$ be the two-tuple address. The connection property of $\mathcal{W}_k$ is delineated by (\ref{eq:conprop1}) and (\ref{eq:conprop2}). As Fig. 7 plots, input wavelength channel

\begin{small}$$X_k=\lubar[19]{{x_n}} \lubar[45]{{x_{n-1}}} \cdots \lubar[19]{{x_2}} \lubar[19]{{x_1}}$$\end{small}or
\begin{small}$$\tilde{X}_k=(\lubar[19]{{x_n}} \lubar[45]{{x_{n-1}}} \cdots \lubar[19]{{x_2}},\lambda_{i})$$\end{small}is connected to output wavelength channel \begin{small}$$Y_k=\lubar[45]{{x_{n-1}}} \cdots \lubar[19]{{x_2}} \lubar[19]{{x_1}}\lubar[19]{{x_n}}$$\end{small}or
\begin{small}$$\tilde{X}_k=(\lubar[45]{{x_{n-1}}} \cdots \lubar[19]{{x_2}} \lubar[19]{{x_1}},\lambda_{i}),$$\end{small}where
\begin{small}$$i=[x_{n}+x_{1}]_{m}.$$\end{small}

Due to the regular topology, we assign the addresses to the input channels and the output channels of $\mathcal{S}(m,n)$ as follows. Firstly, the input channels of $\mathcal{S}(m,n)$ are the input channels of $\mathcal{W}_0$. Therefore, the address of the input channels of $\mathcal{S}(m,n)$ can be the same with that of the input channel of $\mathcal{W}_0$. That is, wavelength channel $\lambda_i$ at the $a$th input of the $p$th input group is assigned with a two-tuple address $\tilde{S}=(\lubar[19]{{s_n}} \lubar[45]{{s_{n-1}}} \cdots \lubar[19]{{s_2}},\lambda_{i})$ and a field address $S=\lubar[19]{{s_n}} \lubar[45]{{s_{n-1}}} \cdots \lubar[19]{{s_2}} \lubar[19]{{s_1}}$, where $s_{n}=p$, $\lubar[45]{{s_{n-1}}} \cdots \lubar[19]{{s_2}}$ is the $m$-ary number of $a$, and $s_{1}=[i-s_{n} ]_m$. The output stage is a column of R-TWC-modules, and can be regarded as the input of a truncated shuffle stage, of which the shuffle network and the column of L-TWC-modules are removed. Thus, we label the output channels in the same way. Different from the input channel address, we use $\tilde{D}=(\lubar[19]{{d_n}} \lubar[45]{{d_{n-1}}} \cdots \lubar[19]{{d_2}},\lambda_i)$ or $D=\lubar[19]{{d_n}} \lubar[45]{{d_{n-1}}} \cdots \lubar[19]{{d_2}} \lubar[19]{{d_1}}$ to denote wavelength channel $\lambda_i$ at the $a$th output of the $p$th output group.

It is clear that $m^{n}\times m^{n}$ AWG-based SEN $\mathcal{S}(m,n)$ has good scalability, since it is constructed from a set of $m\times m$ AWGs and TWCs with a conversion range of $m$. Again, if we keep $m\leq 32$, the coherent crosstalk at each stage can be smaller than 2 dB such that it can be compensated at the outputs of each $m\times m$ AWGs.

This architecture has another important scalability advantage. The most common implementation of TWC is by an opto-electronic transponder that converts optical signals into electronic form and reconverts back into the optical domain onto another wavelength. In doing this operation, the TWC also performs full digital regeneration (3R) of the signals. It should be noted that each shuffle stage is connected to the next by an array of TWCs. Thus, this implies that all the signals are 3R-regenerated from stage to stage. Therefore, signal impairments, such as coherent crosstalk, do not accumulate from stage to stage, making the network highly scalable in terms of number of cascaded stages.

\begin{table*}[ht]
  \centering
  \caption{Routing Path of $R(\tilde{S},\tilde{D})$ in AWG-Based SEN}\label{table:route-path}
  \begin{tabular}{|c|r@{}c@{}c@{}c|c|}
  \hline
  shuffle & \multicolumn{4}{c|}{routing\ path} & wavelength assignment \\
  \hline
  \multirow{2}*{stage 0} &\multirow{2}*{\ }& $\tilde{S}=\tilde{X}_0=(\lubar[19]{{s_n}}\lubar[45]{{s_{n-1}}}\cdots \lubar[19]{{s_2}},\lambda_{i_0}$)& \multirow{2}*{$\rightarrow $} &$\tilde{Y}_0=(\lubar[45]{{s_{n-1}}}\lubar[45]{{s_{n-2}}}\cdots \lubar[19]{{s_1}},\lambda_{i_0})$\ \ \ \ & \multirow{2}*{$i_{0}=[s_{n}+s_{1}]_m$}\\
  & & $S=X_0=\lubar[19]{{s_n}}\lubar[45]{{s_{n-1}}}\cdots \lubar[19]{{s_2}}\lubar[19]{{s_1}}$ & & $Y_0=\lubar[45]{{s_{n-1}}}\lubar[45]{{s_{n-2}}}\cdots \lubar[19]{{s_1}}\lubar[19]{{s_n}}$ \ \ \ \ \ \ \   &\\
  \hline
  \multicolumn{6}{|c|}{ $\cdots$}\\
  \hline
  \multirow{2}*{stage $k$} & \multirow{2}*{$\rightarrow $} &$\tilde{X}_k=(\lubar[45]{{s_{n-k}}} \cdots \lubar[71]{{d_{n-k+2}}},\lambda_{i_k})$& \multirow{2}*{$\rightarrow $} & $\tilde{Y}_k=(\lubar[66]{{s_{n-k-1}}}\cdots \lubar[66]{{d_{n-k+1}}},\lambda_{i_k})$ & \multirow{2}*{$i_{k}=[s_{n-k}+d_{n-k+1}]_m$}\\
  && $X_k=\lubar[45]{{s_{n-k}}}\cdots \lubar[66]{{d_{n-k+2}}}\lubar[66]{{d_{n-k+1}}}$&  &$ Y_k=\lubar[66]{{s_{n-k-1}}}\cdots\lubar[66]{{d_{n-k+1}}} \lubar[45]{{s_{n-k}}}$ &\\
  \hline
  \multicolumn{6}{|c|}{ $\cdots$}\\
  \hline
  output stage & \multicolumn{4}{c|}{
  \begin{tabular}{r@{}l}
  \multirow{2}*{$\rightarrow $} &$(\lubar[19]{{d_n}}\lubar[45]{{d_{n-1}}}\cdots \lubar[19]{{d_2}},\lambda_{i_n})=\tilde{D}$ \\
  &\ \ \  $\lubar[19]{{d_n}}\lubar[45]{{d_{n-1}}}\cdots \lubar[19]{{d_2}}\lubar[19]{{d_1}}=D$
  \end{tabular}
  } & $i_{n}=[d_{n}+d_{1}]_m$\\
  \hline
  \end{tabular}
\end{table*}

\setcounter{equation}{16}
\section{RWA in AWG-based WDM SENs}\label{section:rwa}
One of the most interesting properties of the classical SEN is the self-routing property. Consider a classical $m^{n}\times m^{n}$ SEN, where each input port or each output port is assigned with an $m$-ary $n$-field address. Once input address $S=\lubar[19]{{s_n}} \lubar[45]{{s_{n-1}}} \cdots \lubar[19]{{s_2}} \lubar[19]{{s_1}}$ and output address $D=\lubar[19]{{d_n}} \lubar[45]{{d_{n-1}}} \cdots \lubar[19]{{d_2}} \lubar[19]{{d_1}}$ of a connection request, denoted by $R(S,D)$, are given, the path of this request is uniquely determined \cite{TTlee:Book2010}. The request passes through the first shuffle network and arrives at its output $ \lubar[45]{{s_{n-1}}} \cdots \lubar[19]{{s_2}} \lubar[19]{{s_1}}\lubar[19]{{s_n}}$. After that, it is exchanged to input $\lubar[45]{{s_{n-1}}} \cdots \lubar[19]{{s_2}} \lubar[19]{{s_1}}\lubar[19]{{d_n}}$ of the second shuffle network according to field $d_n$. By the analogy, the complete path is given below:

\begin{small}\begin{equation*}
{\begin{split}
S=\lubar[19]{{s_n}}\lubar[41]{{s_{n-1}}}\cdots&\lubar[19]{{s_1}} \\
\xrightarrow[]{\begin{scriptsize}\mbox{shuffle}\end{scriptsize}} \lubar[38]{{s_{n-1}}}\cdots \lubar[17]{{s_1}}&\lubar[17]{{s_n}}  \xrightarrow[]{\begin{scriptsize}\mbox{exchange}\end{scriptsize}} \lubar[38]{{s_{n-1}}}\cdots \lubar[17]{{s_1}}\lubar[17]{{d_n}}\\
&\cdots \\
\xrightarrow[]{\begin{scriptsize}\mbox{shuffle}\end{scriptsize}} \lubar[63]{{s_{n-k-1}}}\cdots&\lubar[63]{{d_{n-k+1}}}\lubar[38]{{s_{n-k}}}  \xrightarrow[]{\begin{scriptsize}\mbox{exchange}\end{scriptsize}} \lubar[63]{{s_{n-k-1}}}\cdots \lubar[63]{{d_{n-k+1}}}\lubar[38]{{d_{n-k}}}\\
&\cdots \\
\xrightarrow[]{\begin{scriptsize}\mbox{shuffle}\end{scriptsize}} \lubar[17]{{d_n}}\lubar[38]{{d_{n-1}}}\cdots&\lubar[17]{{d_2}}\lubar[17]{{s_1}}  \xrightarrow[]{\begin{scriptsize}\mbox{exchange}\end{scriptsize}} \lubar[17]{{d_n}}\lubar[38]{{d_{n-1}}}\cdots \lubar[17]{{d_2}}\lubar[17]{{d_1}}=D.
\end{split}}
\end{equation*}\end{small}

\begin{figure*}[ht]
\centering
\includegraphics[scale=0.75]{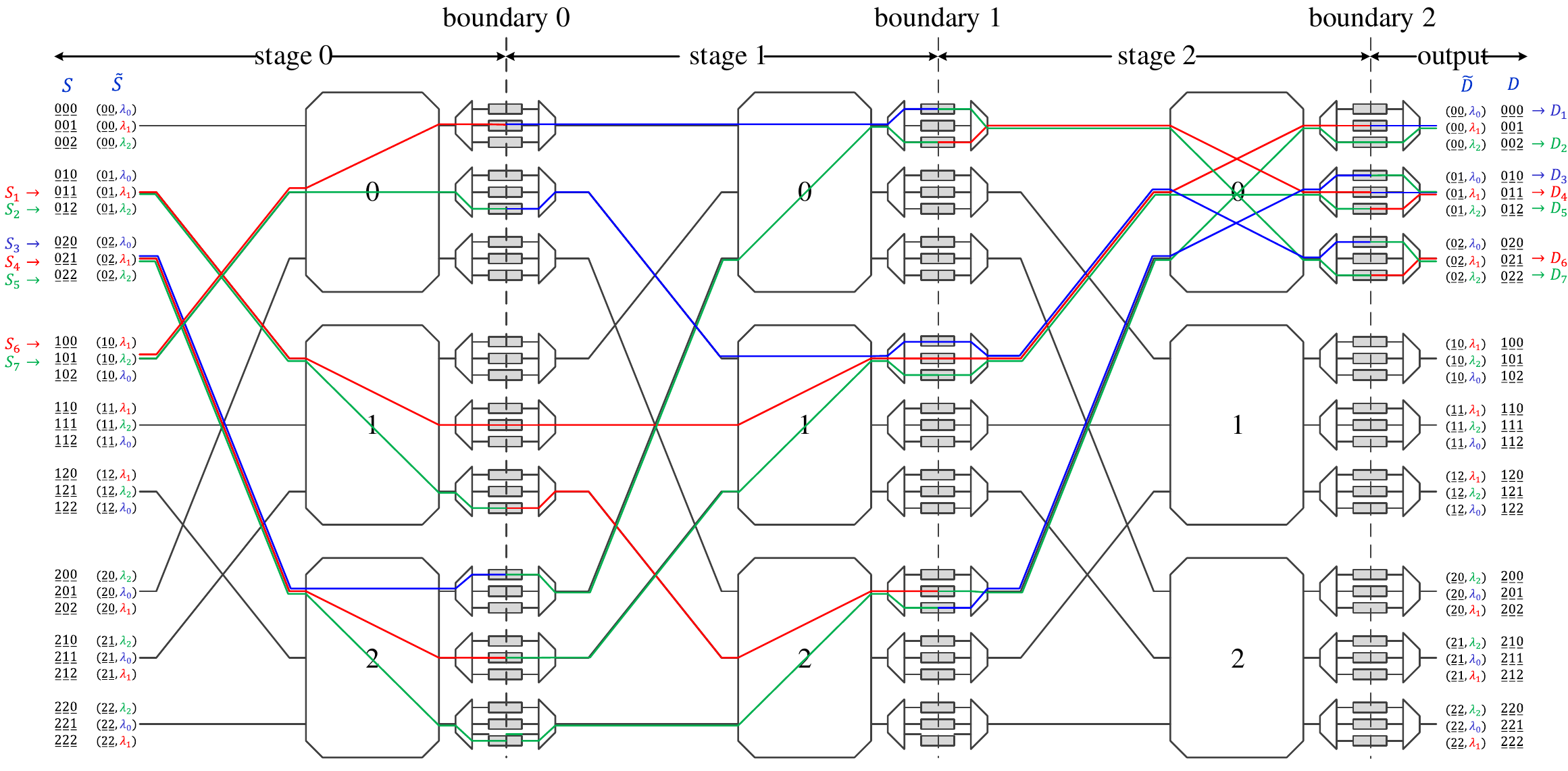}
\caption{Route of multi-requests without contention in $\mathcal{S}(3,3)$.}
\label{fig8}
\end{figure*}

Also, for a set of requests $\pi =\{R_1,R_2,\cdots,R_l\}$ arriving at the network, where $l\leq m^{n}$, the routing is nonblocking if $\pi$ satisfies the following two conditions \cite{TTlee:Book2010}:
\begin{enumerate}[{1)}]
\item
Monotonic: for $S_j > S_i$ if $j> i$, there is $D_1 < D_2 < \cdots < D_n$ or $D_1 > D_2 > \cdots > D_n$.
\item
Concentrated: any input $S_k$ between two active inputs $S_i$ and $S_j$ is active.
\end{enumerate}
\begin{figure*}[!hb]
\begin{footnotesize}
\hrulefill
\begin{equation}
R_1:
{\begin{split}
\tilde{S}=(\ubar{0}\ubar{1}&,\lambda_1) \\
S=\ubar{0}&\ubar{1}\ubar{1}
\end{split}}
\xrightarrow[]{\mathcal{W}_0}
{\begin{split}
(\ubar{1}\ubar{1}&,\lambda_1) \\
\ubar{1}&\ubar{1}\ubar{0}
\end{split}}
\xrightarrow[]{\mbox{boundary\ 0}}
{\begin{split}
(\ubar{1}\ubar{1}&,\lambda_1) \\
\ubar{1}&\ubar{1}\ubar{0}
\end{split}}\xrightarrow[]{\mathcal{W}_1}
{\begin{split}
(\ubar{1}\ubar{0}&,\lambda_1) \\
\ubar{1}&\ubar{0}\ubar{1}
\end{split}}
\xrightarrow[]{\mbox{boundary\ 1}}
{\begin{split}
(\ubar{\bm{1}}\ubar{\bm{0}}&,\bm{\lambda_1}) \\
\ubar{\bm{1}}&\ubar{\bm{0}}\ubar{\bm{0}}
\end{split}}\xrightarrow[]{\bm{\mathcal{W}_2}}
{\begin{split}
(\ubar{\bm{0}}\ubar{\bm{0}}&,\bm{\lambda_1}) \\
\ubar{\bm{0}}&\ubar{\bm{0}}\ubar{\bm{1}}
\end{split}}
\xrightarrow[]{\mbox{boundary\ 2}}
{\begin{split}
(\ubar{0}\ubar{0},&\lambda_0)=\tilde{D} \\
\ubar{0}\ubar{0}&\ubar{0}=D
\end{split}}
\label{eq:self-route2}
\end{equation}
\begin{equation}
R_2:
{\begin{split}
\tilde{S}=(\ubar{1}\ubar{0}&,\lambda_2) \\
S=\ubar{1}&\ubar{0}\ubar{1}
\end{split}}
\xrightarrow[]{\mathcal{W}_0}
{\begin{split}
(\ubar{0}\ubar{1}&,\lambda_2) \\
\ubar{0}&\ubar{1}\ubar{1}
\end{split}}
\xrightarrow[]{\mbox{boundary\ 0}}
{\begin{split}
(\ubar{0}\ubar{1}&,\lambda_0) \\
\ubar{0}&\ubar{1}\ubar{0}
\end{split}}\xrightarrow[]{\mathcal{W}_1}
{\begin{split}
(\ubar{1}\ubar{0}&,\lambda_0) \\
\ubar{1}&\ubar{0}\ubar{0}
\end{split}}
\xrightarrow[]{\mbox{boundary\ 1}}
{\begin{split}
(\ubar{\bm{1}}\ubar{\bm{0}}&,\bm{\lambda_1}) \\
\ubar{\bm{1}}&\ubar{\bm{0}}\ubar{\bm{0}}
\end{split}}\xrightarrow[]{\bm{\mathcal{W}_2}}
{\begin{split}
(\ubar{\bm{0}}\ubar{\bm{0}}&,\bm{\lambda_1}) \\
\ubar{\bm{0}}&\ubar{\bm{0}}\ubar{\bm{1}}
\end{split}}
\xrightarrow[]{\mbox{boundary\ 2}}
{\begin{split}
(\ubar{0}\ubar{0},&\lambda_2)=\tilde{D} \\
\ubar{0}\ubar{0}&\ubar{2}=D
\end{split}}
\label{eq:self-route3}
\end{equation}
\end{footnotesize}
\end{figure*}

As a kind of WDM switching network, the process that $\mathcal{S}(m,n)$ sets up connections for the requests possesses its own feature. Both routing assignment and wavelength assignment should be taken into consideration during connection establishment, which is known as routing and wavelength assignment (RWA) problem. In such kind of network, there is a fundamental condition for contention-free RWA \cite{Ye:TON2015}: if two connections share the same link, they must use different wavelengths; otherwise, they must be link-disjoint. In other words, there is a contention if two connections use the same wavelength at the same link. Though there exist differences between the nonblocking routing in classical SENs and the contention-free RWA in AWG-based WDM SENs, we show in this section that the self-routing property and nonblocking conditions still hold in such WDM SENs.

\subsection{Self-routing property of $\mathcal{W}(m,n)$}
To illustrate the self-routing property of $\mathcal{S}(m,n)$, we take connection $R(\ubar{0}\ubar{1}\ubar{0},\ubar{1}\ubar{1}\ubar{1})$ in Fig. \ref{fig6a} as an example. The two-tuple address of $S=\ubar{0}\ubar{1}\ubar{0}$ is $\tilde{S}=(\ubar{0}\ubar{1},\lambda_0)$, and that of $D=\ubar{1}\ubar{1}\ubar{1}$ is $\tilde{D}=(\ubar{1}\ubar{1},\lambda_2)$. According to the connection property of $\mathcal{W}_{0}(3,3^{2})$, $S$ links to
$$Y_0=\lubar[19]{{s_1}}\lubar[19]{{s_0}}\lubar[19]{{s_2}}=\ubar{1}\ubar{0}\ubar{0}$$
or
$$\tilde{Y}_{0}=(\lubar[19]{{s_1}}\lubar[19]{{s_0}},\lambda_{[s_0+s_2 ]_3})=(\ubar{1}\ubar{0},\lambda_0).$$
Following the self-routing property, wavelength boundary $0$ switches $Y_0$ to
$$X_1=\lubar[19]{{s_1}}\lubar[19]{{s_0}}\lubar[19]{{d_2}}=\ubar{1}\ubar{0}\ubar{1}$$
or $$\tilde{X}_1=(\lubar[19]{{s_1}}\lubar[19]{{s_0}},\lambda_{[s_{0}+d_{2}]_{3}})=(\ubar{1}\ubar{0},\lambda_1),$$
meaning that boundary $0$ changes the wavelength from $\lambda_0$ to $\lambda_1$. By analogy, $R$ passes through the wavelength channels along the path given by (\ref{eq:self-route1}).


It is shown by (\ref{eq:self-route1}) that $R$ passes through a wavelength boundary using one field of $D$, and finally reaches the destination wavelength channel $D$. This example clearly demonstrates that $\mathcal{S}(m,n)$ also possesses the self-routing property.

Table \ref{table:route-path} gives the routing process of $R$ in general case. In particular, at the $k$th shuffle stage $\mathcal{W}_k$, connection $R$ travels from

$$X_k=\lubar[41]{{s_{n-k}}}\cdots \lubar[66]{{d_{n-k+2}}}\lubar[66]{{d_{n-k+1}}}$$or
$$\tilde{X}_k=(\lubar[41]{{s_{n-k}}}\cdots \lubar[71]{{d_{n-k+2}}},\lambda_{i_{k}})$$to
$$Y_k=\lubar[66]{{s_{n-k-1}}}\cdots \lubar[66]{{d_{n-k+1}}}\lubar[41]{{s_{n-k}}}$$or
$$\tilde{Y}_k=(\lubar[66]{{s_{n-k-1}}}\cdots \lubar[71]{{d_{n-k+1}}},\lambda_{i_{k}}),$$where
$$i_k=[s_{n-k}+d_{n-k+1}]_{m},$$and the wavelength of $R$ is converted from
$$\lambda_{i_k}=\lambda_{[s_{n-k}+d_{n-k+1}]_m}$$to
$$\lambda_{i_{k+1}}=\lambda_{[s_{n-k-1}+d_{n-k}]_m}$$at the $k$th wavelength exchange boundary.


\subsection{Nonblocking Self-Routing Conditions}
If there exist multiple requests in $\mathcal{S}(m,n)$ at the same time, wavelength contentions may happen. To see this point, suppose there are two connection requests $R_1(\ubar{0}\ubar{1}\ubar{1},\ubar{0}\ubar{0}\ubar{0})$ and $R_2(\ubar{1}\ubar{0}\ubar{1},\ubar{0}\ubar{0}\ubar{2})$ arriving at $\tilde{S}(3,3)$. According to the self-routing property, the RWAs of $R_1$ and $R_2$ will respectively be (\ref{eq:self-route2}) and (\ref{eq:self-route3}).


In other words, they will have wavelength contention at the input of shuffle stage $\mathcal{W}_2$, since their wavelengths will both be converted to $\lambda_1$ at boundary $1$. From this example, we conclude that:
\begin{enumerate}[{1)}]
\item
Wavelength contention may happen if there exist multiple requests;
\item
A contention may happen at an input of a shuffle stage.
\end{enumerate}
	
The contention will not happen at an output of a shuffle stage if two requests do not use the same wavelength at the same input of a shuffle stage, since each shuffle stage is wavelength contention-free as Property \ref{prop2} in Section \ref{section:modular} shows. This implies that the RWA will be contention-free if any two requests do not use the same wavelength at an input of a shuffle stage.

In the following, we show that the RWA can be contention-free if the set of requests $\pi$ satisfies the monotonic and concentrated conditions.
\newtheorem{theorem}{Theorem}
\begin{theorem}
AWG-based WDM SEN $\mathcal{S}(m,n)$ is contention-free if the requests in this network are monotonic and concentrated.
\end{theorem}
\begin{IEEEproof}
Consider two requests $R(S,D)$ and $R'(S',D')$, where

\begin{equation}{
S'=\lubar[19]{{s_{n}'}}\cdots\lubar[19]{{s_{2}'}}\lubar[19]{{s_{1}'}} > S=\lubar[19]{{s_{n}}}\cdots\lubar[19]{{s_{2}}}\lubar[19]{{s_{1}}}
}\end{equation}and

\begin{equation}{
D'=\lubar[19]{{d_{n}'}}\cdots\lubar[19]{{d_{2}'}}\lubar[19]{{d_{1}'}} > D=\lubar[19]{{d_{n}}}\cdots\lubar[19]{{d_{2}}}\lubar[19]{{d_{1}}}.
}\end{equation}
The number of input channels between $S$ and $S'$ is $|S'-S+1|$, and that of output channels between $D$ and $D'$ is $|D'-D+1|$. It follows from the concentrated condition

\begin{equation}{
|S'-S| \leq |D'-D|.
\label{theorem1assume}
}\end{equation}
According to the self-routing property, $R$ and $R'$ respectively reach the output wavelength channel of $\mathcal{W}_k$ as follows:

\begin{equation}{
Y_k=\lubar[63]{{s_{n-k-1}}}\cdots \lubar[63]{{d_{n-k+1}}} \lubar[41]{{s_{n-k}}} }\end{equation}
or

\begin{equation}{
\tilde{Y}_{k}=(\lubar[63]{{s_{n-k-1}}}\cdots \lubar[63]{{d_{n-k+1}}},\lambda_i),
\label{theorem1tag1}
}\end{equation}
and

\begin{equation}{
Y'_k=\lubar[63]{{s'_{n-k-1}}}\cdots \lubar[63]{{d'_{n-k+1}}} \lubar[41]{{s'_{n-k}}}}\end{equation}
or

\begin{equation}{
\tilde{Y}'_{k}=(\lubar[63]{{s'_{n-k-1}}}\cdots \lubar[63]{{d'_{n-k+1}}},\lambda_{i'}),
}\end{equation}
where

\begin{equation}{
i=[s_{n-k}+d_{n-k+1}]_m
}\end{equation}
and

\begin{equation}{
i'=[s'_{n-k}+d'_{n-k+1}]_m.
}\end{equation}Assume that $R$ and $R'$ compete for the same input wavelength channel of $\mathcal{W}_{k+1}$. In this case, they first reach different output wavelength channels of $\mathcal{W}_k$, which means

\begin{equation}{
\lubar[68]{{s_{n-k-1}}}\cdots \lubar[68]{{d_{n-k+1}}}=\lubar[68]{{s'_{n-k-1}}}\cdots \lubar[68]{{d'_{n-k+1}}}
\label{theorem1tag1}
}\end{equation}
and

\begin{equation}{
s_{n-k} \not= s'_{n-k}.
}\end{equation}
According to $d_{n-k}$ and $d'_{n-k}$, boundary $k$ then converts $R$ and $R'$ respectively to the same input wavelength channel of $\mathcal{W}_{k+1}$, i.e.,

\begin{align}{
\tilde{X}_{k+1}&=(\lubar[68]{{s_{n-k-1}}}\cdots \lubar[68]{{d_{n-k+1}}},\lambda_j)\nonumber\\
&=(\lubar[68]{{s'_{n-k-1}}}\cdots \lubar[68]{{d'_{n-k+1}}},\lambda_j)\nonumber\\
&=\tilde{X}'_{k+1},
}\end{align}
where $\lambda_j$ denotes the wavelength of input channel. Thus, we have

\begin{equation}{
d_{n-k} =[j-s_{n-k-1}]_m=[j-s'_{n-k-1}]_m = d'_{n-k}.
\label{theorem1tag2}
}\end{equation}According to (\ref{theorem1tag1})-(\ref{theorem1tag2}), we have
$$\begin{small}
{\begin{split}
|S'-S|&=|\lubar[19]{{s'_n}}\cdots \lubar[19]{{s'_1}}-\lubar[19]{{s_n}}\cdots \lubar[19]{{s_1}}| \\
&=|\lubar[19]{{s'_n}}\cdots \lubar[45]{{s'_{n-k}}}\ubar{0}\cdots \ubar{0}|-|\lubar[19]{{s_n}}\cdots \lubar[45]{{s_{n-k}}}\ubar{0}\cdots \ubar{0}| \\
&=|\lubar[19]{{s'_n}}\cdots \lubar[45]{{s'_{n-k}}}-\lubar[19]{{s_n}}\cdots \lubar[45]{{s_{n-k}}}|\times m^{n-k-1}\\
&\geq m^{n-k-1}
\end{split}}\end{small}
$$
and
$$\begin{small}
{\begin{split}
|D'-D|&=|\lubar[19]{{d'_n}}\cdots \lubar[19]{{d'_1}}|-|\lubar[19]{{d_n}}\cdots \lubar[19]{{d_1}}| \\
&=|\ubar{0}\cdots \ubar{0} \lubar[68]{{d'_{n-k-1}}}\cdots \lubar[19]{{d'_1}}|-|\ubar{0}\cdots \ubar{0} \lubar[68]{{d_{n-k-1}}}\cdots \lubar[19]{{d_1}}| \\
&\leq m^{n-k-1} - 1,
\end{split}}\end{small}
$$
which indicates
\begin{equation}{
|S'-S| > |D'-D|.
}\end{equation}This contradicts (\ref{theorem1assume}). Therefore, $R$ and $R'$ must not share the same input wavelength channel of $\mathcal{W}_{k+1}$.

As the two requests are arbitrarily chosen, the RWAs in the AWG-based WDM SEN are contention-free.
\end{IEEEproof}
Fig. 8 gives an example where the set of requests is as follows:

\begin{small}\begin{equation}{
\begin{array}{c}
\pi = \{R_{1}(\ubar{0}\ubar{1}\ubar{1},\ubar{0}\ubar{0}\ubar{0}),R_{2}(\ubar{0}\ubar{1}\ubar{2},\ubar{0}\ubar{0}\ubar{2})
R_{3}(\ubar{0}\ubar{2}\ubar{0},\ubar{0}\ubar{1}\ubar{0}),R_{4}(\ubar{0}\ubar{2}\ubar{1},\ubar{0}\ubar{1}\ubar{1}),\\
R_{5}(\ubar{0}\ubar{2}\ubar{2},\ubar{0}\ubar{1}\ubar{2}),R_{6}(\ubar{1}\ubar{0}\ubar{0},\ubar{0}\ubar{2}\ubar{1}),R_{7}(\ubar{1}\ubar{0}\ubar{1},\ubar{0}\ubar{2}\ubar{2})\}.
\end{array}
}\end{equation}\end{small}It is clear that $\pi$ satisfies the monotonic and concentrated conditions. We can see that there is no contention even if some requests might share the same fiber link. An example is that $R_1$, $R_4$, and $R_7$ at an input of $\mathcal{W}_2$ in Fig. \ref{fig8} use different wavelengths.

Let's consider an extreme case where all the $m^{n}$ input wavelength channels of $\mathcal{S}(m,n)$ are busy. Recall that there are $m^n$ TWCs at each exchange stage and the shuffle network at each shuffle stage contains $m^n$ wavelength channels. Therefore, the network will achieve 100\% utilization in this case.

\section{Conclusion}\label{section:conclu}
In this paper, we propose a method to construct an AWG-based optical SEN. We first study the equivalence between a single AWG and a classical shuffle network, based on which we devise a systematic approach to design a large-scale AWG-based WDM shuffle network. Combining the AWG-based WDM shuffle networks and the TWC-modules, we obtain an AWG-based WDM SEN, which is scalable due to the following reasons. First, the wavelength granularity, the coherent crosstalk, and the conversion range of the TWCs in the network are small since the network only employs a set of small-size AWGs associated with the same wavelength set. Second, the cabling complexity at each shuffle stage is low. Third, the RWA in this network is consistent with that in classical SENs. Fourth, the network can achieve 100\% utilization if the input wavelength channels are all busy. Fifth, signals are 3R regenerated from stage to stage, ensuring high scalability in terms of number of stages.

\ifCLASSOPTIONcaptionsoff
\newpage
\fi

\bibliography{IEEEabrv,refDJJ}


\end{document}